\documentclass[preprint,nofootinbib,floatfix]{revtex4-2}
\usepackage{graphicx}
\usepackage{epsfig}
\usepackage{epstopdf}
\usepackage{amsfonts}
\usepackage{amssymb}
\usepackage{amsbsy}
\usepackage{amsmath}
\usepackage{mathrsfs}
\usepackage{latexsym}
\usepackage{natbib}
\usepackage{bm}
\usepackage{color}
\usepackage{braket}
\usepackage{slashed}
\usepackage{pgfplots}
\usepackage{graphicx,subcaption,lipsum}
\usepackage[normalem]{ulem}

\usepackage{tikz}
\usetikzlibrary{shapes,arrows,shadows}
\usepackage{hyperref,url}
\hypersetup{
	colorlinks=true,       
	linkcolor=blue,          
	citecolor=blue,        
	filecolor=blue,      
	urlcolor=blue           
}

\begin{document}

\affiliation{ Department of Physics, 
Indian Institute of Technology Bombay,
Mumbai - 400076, India }


\date{\today}

\title{From Horndeski action to the Callan-Giddings-Harvey-Strominger model and beyond}

%
\author{Susobhan Mandal}
\email{sm12ms085@gmail.com}
\thanks{Equal contribution to this work.}
\affiliation{Department of Physics, Indian Institute of Technology Bombay, Mumbai 400076, India}
\author{Tausif Parvez}
\email{214120002@iitb.ac.in}
\thanks{Equal contribution to this work.}
\affiliation{Department of Physics, Indian Institute of Technology Bombay, Mumbai 400076, India}
\author{S. Shankaranarayanan}
\email{shanki@iitb.ac.in}
\affiliation{Department of Physics, Indian Institute of Technology Bombay, Mumbai 400076, India}	%

\begin{abstract}
The knowledge of what entered black hole (BH) is completely lost as it evaporates. This contradicts the unitarity principle of quantum mechanics and is referred to as the information loss paradox. Understanding the end stages of BH evaporation is key to resolving this paradox. As a first step, we need to have exact models that can mimic 4-D BHs in General relativity in classical limit and have a systematic way to include high-energy corrections. While there are various models in the literature, there is no systematic procedure by which one can study high-energy corrections. In this work, for the first time, we obtain Callan, Giddings, Harvey, and Strominger (CGHS) --- a (1+1)-D --- model from 4-D Horndeski action --- the most general scalar-tensor theory that does not lead to Ostrogradsky ghosts. We then show that 4-D Horndeski action can systematically provide a route to include higher-derivative terms relevant at the end stages of black hole evaporation. We derive the leading order Hawking flux while discussing some intriguing characteristics of the corrected CGHS models. We compare our results with other works and discuss the implications for primordial BHs.
\end{abstract}

\maketitle
\newpage

\section{Introduction}

The observation of the merging black holes (BHs) by LIGO-VIRGO served as an extraordinary validation of the tenets of general relativity (GR). However, it remains unclear whether BHs adhere to the same principles of quantum mechanics (QM) as all other known objects~\cite{Peres:2002wx,Bartlett:2006tzx}. Classically, BHs are perfect absorbers but cannot emit anything; their physical temperature is absolute zero, and their entropy is
infinite~\cite{2001-Wald-LRR,Page:2004xp,Das:2008sy,Carlip:2014pma,Chen:2014jwq}. Nevertheless, once QM is applied, BH evaporates by emitting Hawking radiation that carries \emph{no information} about their microstates~\cite{hawking1975particle}. This conundrum known as \emph{BH information paradox} has remained unresolved for four decades~\cite{hawking1976breakdown,giddings1992black, harvey1992quantum, preskill1992black, page1994black,giddings1994quantum, strominger1995houches,mathur2009exactly,hossenfelder2010conservative,
harlow2016jerusalem,Raju:2020smc,2022-Calmet-Hsu-EPL}.

Most BHs detected by LIGO-VIRGO are heavier than the previously known population of stellar-mass BHs that were indirectly inferred from EM observations~\cite{2019-GWTC1-PRX,2021-GWTC3-Arxiv}. 
The anomalously heavy BHs detected by LIGO have 
renewed interest in Primordial Black Holes (PBHs)~\cite{2016-Sasaki.etal-PRL,2016-Bird.etal-PRL,2016-Hayasaki.etal-PASJ}.
Current constraints suggest the PBHs in mass windows $10^{17} - 10^{23}~{\rm gm}$ are potential dark matter candidates~\cite{Carr:2020gox,2021-Carr-Kuhnel-ARNPS,2021-Menaetal-Frontiers,2021-Green.Kavanagh-JPG}. Interestingly, Hawking radiation is significant for these masses, and Hawking temperature~\cite{hawking1975particle,1976-Page-PRD,1978-Sanchez-PRD}: 
\begin{equation}
T_{\rm H} = 
\left(\frac{\hbar c^3}{G k_{_{B}}}\right) \frac{1}{8 \pi M} \sim 10^{-7} \left(\frac{M_{\odot}}{M} \right) \ K
\end{equation} 
is large. Note $k_B$($G$) is Boltzmann (4-D Newton's) constant, and $M$ is the BH mass. Higher-order curvature terms must be considered since smaller masses experience substantial tidal forces. For Schwarzschild BH, this is evident from Kretschmann scalar:
\begin{equation}
R_{\mu\nu\rho\sigma}R^{\mu\nu\rho\sigma} \propto (G M)^{-4} \, .
\end{equation}
From the above two expressions, we see that as the BH mass decreases due to evaporation, the Hawking temperature and the strength of the Kretschmann scalar near the horizon increase rapidly. This clearly shows that the evaporation of small mass PBHs whose mass ranges from $10^{-15} - 10^{-5}M_{\odot}$ necessitates including higher-order curvature terms and hence requires one to go beyond GR~\cite{2022-Shanki.Joseph-GRG}. The accelerated emission process for smaller mass BHs leads to a cataclysmic release of radiation, annihilating the BHs~\cite{hawking1975particle,1976-Page-PRD,1978-Sanchez-PRD}.

The evolution of radiating BH requires the theory of quantum gravity, which remains elusive. Due to the non-linear nature of gravity, semiclassical gravity fails at the Planck length scale or involving singularities~\cite{1984-Birrell.Davies-Book,2009-Parker-Toms-Book}. 
Hence, accurate characterization of the ultimate phase of BH evaporation within a semiclassical framework is unattainable. 
However, the situation is not entirely pessimistic; \emph{a glimmer of hope exists}. Numerical results of 4-D BHs show around $90\%$ of Hawking radiation is in $s$-waves~\cite{1976-Page-PRD,1978-Sanchez-PRD}. Thus, restricting to $s$-waves captures almost all the essential physics of BH radiation~\cite{Visser:2001kq,2003-Shanki-PRD}. 

For 4-D spherically symmetric space-times, $s$-wave corresponds to picking out the $t-r$ plane and ignoring (or integrating) the two angular degrees of freedom --- \emph{effective 2-D gravity}. However, 2-D gravity must be distinct from GR since the Einstein tensor is topologically trivial. The simplest possible modification to obtain dynamics in 2-D is via direct coupling of the Ricci scalar and scalar (dilaton) field~\cite{DHoker:1982wmk,Teitelboim:1983ux,Jackiw:1984je,callan1992evanescent,Mann:1991md,russo1992black,McGuigan:1991qp,1992-Frolov-PRD,Kazakov:2001pj,2002-Grumiller.etal-PRep,Mertens:2022irh,Nejati:2023hpe}. Such a 2-D gravity can be obtained by either treating the system as $D = 4$ and imposing spherical symmetry in the equations of motion~\cite{1994-Kuchar-PRD} or 
imposing spherical symmetry in the action~\cite{1972-Berger.etal-PRD,DHoker:1982wmk,Teitelboim:1983ux}. Both approaches are classically equivalent. 

Given this, we ask: Can we study the modifications to Hawking radiation in $s-$waves due to the higher-order curvature corrections? If yes, what modifications must be included, and how can the comparison be made? In this work, we address these issues within the framework of the CGHS model~\cite{callan1992evanescent,Strominger:1994tn} described by the action: 
\begin{equation}\label{CGHS action}
S_{\text{CGHS}} = \int \frac{d^{2} x}{4\pi} \sqrt{-g} \ e^{-2 \varphi}\Big[\mathcal{R} + 
4\nabla_{\alpha}\varphi\nabla^{\alpha}\varphi + 4\lambda^{2}\Big] \, ,
\end{equation}
where, $\varphi$ is the dilaton field, $\lambda^2$ is \emph{cosmological constant}, and $\mathcal{R}$ is the 2-D Ricci scalar. 
The above action has an an exact classical BH solution and hence, 
is a useful toy model for studying BH thermodynamics. The standard calculations reveal Hawking radiation with a temperature of $\lambda/{2 \pi}$. It was shown that the effective action including the backreaction of the Hawking radiation gives the evaporating BH analogy~\cite{russo1992black}. 

However, the CGHS action \eqref{CGHS action} can not be derived from a 4-D gravity action.  This has been a key obstacle in generalizing the CGHS model by including higher-order curvature corrections and obtaining corrections to Hawking radiation at the late stages of evolution. \emph{This work fills this void by obtaining the above CGHS action from 4-D Horndeski gravity}~\cite{1974-Horndeski-IJTP,Clifton:2011jh,Kobayashi:2019hrl}. We explicitly show that the spherical reduction of the lowest order terms in 4-D Horndeski action identically leads to the above action. We then systematically obtain the higher curvature corrections to CGHS action and study the implications. In particular, we study two models to discuss the qualitative features of Hawking radiation and compare them with the earlier results~\cite{jacobson1991black, unruh1995sonic}. 

\noindent \section{4-D gravity to CGHS:} Let us consider the following Horndeski action~\cite{1974-Horndeski-IJTP,Clifton:2011jh,Kobayashi:2019hrl}.
\begin{eqnarray}
\label{eq:Horndeskiaction}
S &=& \frac{1}{16 \pi G} \int d^4x \sqrt{-g^{(4)}} \big[G_{2}(X,\Phi)
+ G_{4}(\Phi) R^{(4)} \nonumber  \\ && 
-G_{3}(X,\Phi)\Box{\Phi}  - G_{5}(\Phi) G^{(4)}_{\mu\nu}\nabla^{\mu} \Phi \nabla^{\nu}\Phi\big]
\end{eqnarray}
where $R^{(4)}$ is the 4-D Ricci scalar corresponding to the metric $g_{\mu\nu}^{(4)}$, $\Phi$ is the mass-dimension zero scalar field, $G_i$'s are functions of $\Phi$\footnote{The last term in Eq.~\eqref{eq:Horndeskiaction} is different compared to Ref.~\cite{Kobayashi:2019hrl} by a boundary term.}, $G_2$ and $G_3$ are also functions of $X \equiv - \frac{1}{2}\nabla_{\mu}\Phi\nabla^{\mu}\Phi$. In the above action, the last two terms contain higher-derivatives. First, we set 
\begin{equation}
\label{eq:Horn-CGHS-condition01}
G_2 = \Omega(\Phi) X -  V(\Phi);  
G_3 = G_5 = 0; 
G_4= \Phi.
\end{equation}
To do spherical reduction, we choose the metric ansatz:
\begin{equation}\label{metric ansatiz}
ds^{2} = g_{AB}dx^{A}dx^{B} + \rho^{2}(\{x^{A}\})(d\theta^{2} + \sin^{2}\theta d\phi^{2}),
\end{equation}
where $(\theta, \phi)$ are the spherical polar coordinates, $\rho$ acts as a dynamical radial coordinate which depends on $x^{A}$ and $A = 0, 1$. For the above line-element, we have:
\begin{equation}\label{spherical reduction}
R^{(4)} = \mathcal{R} + 2 \left(1 +  \nabla_{\mu}\rho\nabla^{\mu}\rho
- \Box\rho^{2} \right)/{\rho^{2}} \, ,
\end{equation}
where the covariant derivative is defined \textit{w.r.t} the 2-D metric $g_{AB}$. Substituting Eq.~\eqref{spherical reduction} in action \eqref{eq:Horndeskiaction}, we get:
\begin{eqnarray}
S & =&  \frac{1}{4 G}\int d^{2}x \sqrt{-g} \left[ \rho^{2} \Phi\mathcal{R} + 
{2\Phi}  + {2\Phi} \nabla_{\mu}\rho\nabla^{\mu}\rho \right.
\nonumber \\
 & -& \left. {2\Phi} \Box\rho^{2} + \Omega(\Phi) X - V(\Phi)\right] \, .
\label{eq:Horn-CGHS01}
\end{eqnarray}
Since the above action has two scalar fields ($\Phi$ and $\rho$), we must 
choose a relation between the two scalars that will lead to CGHS action \eqref{CGHS action}. Substituting 
\begin{equation}
\label{eq:Horn-CGHS-condition02}
\Omega(\Phi) = -{6}/{\Phi}; 
{V(\Phi)} = A\Phi + B\Phi^{3}; \Phi = {\Lambda}/{\rho}, 
\end{equation}
in action \eqref{eq:Horn-CGHS01}, we have:
  \begin{eqnarray}
\!\!\! S = \!\! \frac{\Lambda}{4 G} \int \!\! d^2x\sqrt{-g} \bigg[ \rho {\cal R} +\frac{2 - B \Lambda^2}{\rho}+ \rho [\nabla \log\rho]^2 - A\rho \bigg]  \nonumber
    \end{eqnarray}
where $\Lambda, A$ and $B$ are dimensionful constants. Redefining, $\rho = \Lambda e^{-2\varphi}$ in the above action leads to:
{\small
\begin{eqnarray}
S = \frac{\Lambda^2}{4 G} \int d^2x\sqrt{-g}\bigg[ e^{-2\varphi} 
\left[ {\cal R} + 4(\nabla\phi)^2 - A\right] + 
\left[\frac{2}{\Lambda^2} - B \right] e^{2 \varphi}  \bigg] \nonumber
   \end{eqnarray} 
   }
Setting
\begin{equation}
A = -  4\lambda^{2}~{\rm and}~B = {2}/{\Lambda^{2}},
\end{equation} 
in the above action leads to the following:
\begin{equation}
S = \left(\Lambda^2 \pi/{G} \right) S_{\rm CGHS}
\label{eq:Horn-CGHSFin}
\end{equation}
This is the first key result of this work, regarding which we want to stress the following points: First, upto an overall constant factor, the above reduced-action is identical to the CGHS action \eqref{CGHS action}. Since $\Lambda$ is an arbitrary constant (not related to the cosmological constant), setting $\Lambda^{2} = G/\pi$, we get CGHS action, including the factors. 
To our knowledge, this is the first time such a mapping has been established. 
Second, the reader might consider this approach to
be contrived. 
As shown in Sec. I in supplemental material \cite{2023-Mandal.etal-Supp}, obtaining the CGHS action from a dimensional-reduced Einstein-Hilbert action in arbitrary dimensions is impossible. The result is also valid for conformally related 4-D spherically symmetric space-times. The analysis can be extended to different topologies like $S^m \times S^n$, ($m, n \in \mathbb{Z}^+$) and it can be shown that the Einstein-Hilbert action can not lead to CGHS. This implies that a pure metric theory of gravity in arbitrary dimensions via dimensional reduction can not lead to CGHS action. As shown above, 
the CGHS action \emph{can only} be obtained from the dimensional reduction of scalar-tensor gravity theory~\cite{1961-Brans.Dicke-Phys.Rev.,2005-Brans-Talk,2010-DeFelice.Tsujikawa-LRR,2010-Sotiriou.Faraoni-RMP,Nojiri:2010wj}, which can be interpreted as a time-varying gravitational “constant” represented by a scalar field $\varphi$.
Third, an attentive reader might identify that the Horndeski action \eqref{eq:Horndeskiaction} in the limit $G_3 = G_5 = 0$ corresponds to Brans-Dicke theory~\cite{1961-Brans.Dicke-Phys.Rev.,2005-Brans-Talk,2010-DeFelice.Tsujikawa-LRR,2010-Sotiriou.Faraoni-RMP,Nojiri:2010wj}. It is interesting to note that CGHS action \eqref{CGHS action} is equivalent to 2-D Brans-Dicke  
theory. To see this, redefining the dilaton field ($\varphi$) as  $ \Psi = e^{-2\varphi}$ in the  CGHS action \eqref{CGHS action}, we get:
\begin{equation}
\! S_{\text{CGHS}} = \frac{1}{4\pi}\int \!\! 
d^{2}x \sqrt{-g} \Big[\Psi\mathcal{R} + \frac{1}{\Psi}
\nabla_{\mu}\Psi\nabla^{\mu}\Psi + 4\lambda^{2}\Psi\Big]
\end{equation}
The above action maps to 2-D Brans-Dicke action exactly for $\Omega(\Psi) = - 1/\Psi$ and $V({\Psi}) = - 4\lambda^{2} \Psi$. 

Fourth, since CGHS action is obtained from 4-D Brans-Dicke theory, the spherically symmetric solutions of Brans-Dicke theory are solutions of the CGHS action~\cite{grumiller2002dilaton}. Also, as shown first by Hawking \cite{hawking1972black} and generalized by Faraoni and Sotiriou~\cite{Sotiriou:2011dz}, a stationary spherically symmetric solution as the end-state of collapse for a large class of scalar-tensor theories of gravity isolated is same as that of general relativity. Hence, the static line-element \eqref{metric ansatiz} is that of GR.
Lastly, the above mapping provides a route to introduce higher-order curvature corrections relevant at the end stages of BH evaporation. It is well-known that higher curvature terms introduce higher-derivative terms, leading to Ostrogradski instability~\cite{1983-Barth-Christensen-PRD,1990-Simon-PRD,2001-Hawking.Hertog-PRD}. However, Horndeski action \eqref{eq:Horndeskiaction} is the 
most general scalar-tensor gravity action that leads to II-order equations of motion and hence does not possess Ostrogradski instability~\cite{1974-Horndeski-IJTP,Clifton:2011jh,Kobayashi:2019hrl}. Thus, the spherical reduction of Horndeski action \eqref{eq:Horndeskiaction} will contain higher curvature corrections without leading to any instability and can provide crucial insights about the late stages of Hawking radiation. As $G_3$ and $G_5$ (in action \eqref{eq:Horndeskiaction}) are unknown, many ways exist to include higher-curvature corrections to CGHS. In the rest of this work, we consider two specific forms that indicate possible effects of higher curvature corrections at the Hawking flux.

\noindent \section{Beyond classical CGHS model:} 
To go beyond CGHS, we need to switch on the coefficients $G_3$ and $G_5$ in the action \eqref{eq:Horndeskiaction}. Specifically, we set 
\begin{equation}
G_{3}(X,\Phi) = X \, f(\Phi); \Omega(\Phi) = - {6}/{\Phi} + {h(\Phi)}/{\Lambda^{2}}, 
\end{equation}
and $G_5$ is an arbitrary function of $\Phi$. 
Using the metric ansatz \eqref{metric ansatiz}, making the spherical reduction
and substituting the relations, $\Phi = \Lambda/\rho, \rho = \Lambda e^{- 2\varphi}$, the action \eqref{eq:Horndeskiaction} reduces to: 
\begin{eqnarray}
 & & S = \frac{\Lambda^2 \pi}{G} \, S_{\rm CGHS} - \frac{\Lambda^2}{4 G} 
\int d^{2}x  \sqrt{-g} \Big[ \frac{4 h(\Phi)}{\Lambda^2} X_{\varphi}  \nonumber \\
& & - 8 e^{2 \varphi} f(\Phi) X_{\varphi}  \left[\Box\varphi - 4 X_{\varphi} \right]
-  G_5(\Phi) \left[ 16 \nabla_{\mu} \nabla_{\nu} \varphi \nabla^{\mu} \varphi \nabla^{\nu} \varphi \right. \nonumber \\
\label{eq:CGHScorrec}
& & \left. + \frac{8}{\Lambda^2} e^{4 \varphi} X_{\varphi}  + 64 X_\varphi^2 + 32 
X_{\varphi} \Box\varphi \right] \Big]
\end{eqnarray}
where $X_{\varphi} \equiv - \frac{1}{2}\nabla_{\mu}\varphi \nabla^{\mu}\varphi$.
Sec. II in Ref.~\cite{2023-Mandal.etal-Supp} contains the detailed derivation of the above action.

As can be seen, the corrections to CGHS contain three unknown coefficients --- $f(\Phi), h(\Phi)$ and $G_5(\Phi)$. We consider two simple forms of these three functions, which leads to two classes of CGHS corrected models:
\begin{eqnarray}
\mbox{\bf Model 1} \quad & & G_5 = \mathcal{G};~h(\Phi) = 2 \Phi^2\mathcal{G};~ f(\Phi) = - 
\frac{6 \mathcal{G}}{\Phi} \nonumber \\
\mbox{\bf Model 2}\quad  & & G_5 = \frac{\mathcal{G}}{\Phi};~h(\Phi) = 2 \Phi\mathcal{G};~ f(\Phi) = - \frac{6 \mathcal{G}}{\Phi^2}~~~~~ 
\end{eqnarray}
where ${\cal G}$ is a constant. Varying the above action w.r.t the dilaton field ($\varphi$) and the 2-D metric, leads to: 
{\small
\begin{eqnarray}
\partial_{\pm}^{2}\varphi & =&  2\partial_{\pm}\omega\partial_{\pm}\varphi - 
2 \epsilon_i G_5[\Phi(\varphi)] e^{2(\varphi - 
\omega)}(\partial_{\pm}\varphi)^{3}\partial_{\mp}\varphi \nonumber \\
%
\partial_{+}\partial_{-}\varphi & =& \frac{\lambda^2 e^{2\omega}}{2}+ 2\partial_{+}\varphi\partial_{-}
\varphi + \epsilon_i G_5[\Phi(\varphi)] e^{2(\varphi - \omega)}(\partial_{+}\varphi\partial_{-}\varphi)^{2} \nonumber \\
\partial_{+}\partial_{-}\omega & = & \frac{\lambda^2}{2}e^{2\omega} + 2\partial_{+}\varphi\partial_{-}
\varphi [1 + \epsilon_i \lambda^2  G_5[\Phi(\varphi)] e^{2\varphi}] \nonumber \\
 & +& C_{i} \epsilon_i G_5[\Phi(\varphi)] e^{2(\varphi - \omega)}(\partial_{+}\varphi\partial_{-}\varphi)^{2},
 \label{eq:EOM-CGHSCorrec}
\end{eqnarray}
}
where $i = 1, 2$ corresponding to the two models, 
\[
C_{1} = 10, C_{2} = 7; \epsilon_1 = 2048 {G\lambda^{2}}/{\mathcal{G}}, 
\epsilon_2 = {2560 G\lambda^{2}}/{\mathcal{G}} \, .
\]
In the above equations, the 2-D line-element is set to be
\begin{equation}
ds^2 =  - e^{2 \omega(x_+,x_-)} dx_+ dx_-  = - e^{2 \tilde{\omega}(\sigma_+,\sigma_-)}
d\sigma_+ d\sigma_-
\label{eq:2Dmetric}
\end{equation}
where $x_{\pm}, \sigma_{\pm}$ are the null coordinates. Secs. III and IV in Ref. \cite{2023-Mandal.etal-Supp} contain the details of the derivation. 

For static configurations, the above expressions can be written in a compact form by rewriting in terms of $X_{\varphi}$ (see Sec. V in Ref.~\cite{2023-Mandal.etal-Supp}):
\begin{eqnarray}
& & \frac{\partial_{\sigma}X_{\varphi}}{ \partial_{\sigma} \varphi} = 1 + 2X_{\varphi} - 
\frac{\epsilon_{i}}{2} \frac{G_{5}(\Phi)\Phi}{\mathcal{G}} X_{\varphi}^{2} ; 
(\partial_{\sigma}\varphi)^{2}  = - 2 e^{2\omega} X_{\varphi} 
\nonumber \\
& & \partial_{\sigma}^{2}\omega  = - 4e^{2\omega}\Big[\frac{1}{2} + X_{\varphi}\left[ 1 + 
\epsilon_{i}\frac{G_{5}(\Phi)\Phi}{\mathcal{G}}\right] \nonumber \\
 & & \qquad \quad + \frac{c_{i}\epsilon_{i}}{4}\left[\frac{G_{5}(\Phi)\Phi}{\mathcal{G}}\right]
 X_{\varphi}^{2}\Big] \, ,
\end{eqnarray}
where $\sigma = (\sigma^{+} - \sigma^{-})/2$ and $\lambda x^{\pm} \rightarrow \pm e^{\pm\lambda\sigma^{\pm}}$~\cite{callan1992evanescent}.  The above equations form a set of coupled differential equations of  $\varphi, X_{\varphi}$, and $\omega$. Before we proceed with the solution, we want to mention the following key points: First, equations of motion of the CGHS and the two models possess shift symmetry --- $\varphi \to \varphi + c_0$ where $c_0$ is a constant. Note that the action \eqref{eq:CGHScorrec} is not invariant under this symmetry. This is analogous to the scaling symmetry (of the coordinates) of a simple harmonic oscillator. This is a symmetry of the EOM of the harmonic oscillator but not of the Lagrangian. 
Second, the shift symmetry corresponds to $\rho \to e^{-2c_0} \rho$. Specifically, the solutions to the above  equations for a given $\rho(x_-,x_+)$ will lead to an infinite number of identical solutions scaled by $e^{2 c_0}$. 
\begin{figure*}[!hbt]
 \begin{subfigure}{.24\textwidth}
  \centering
    \includegraphics[width=1\linewidth]{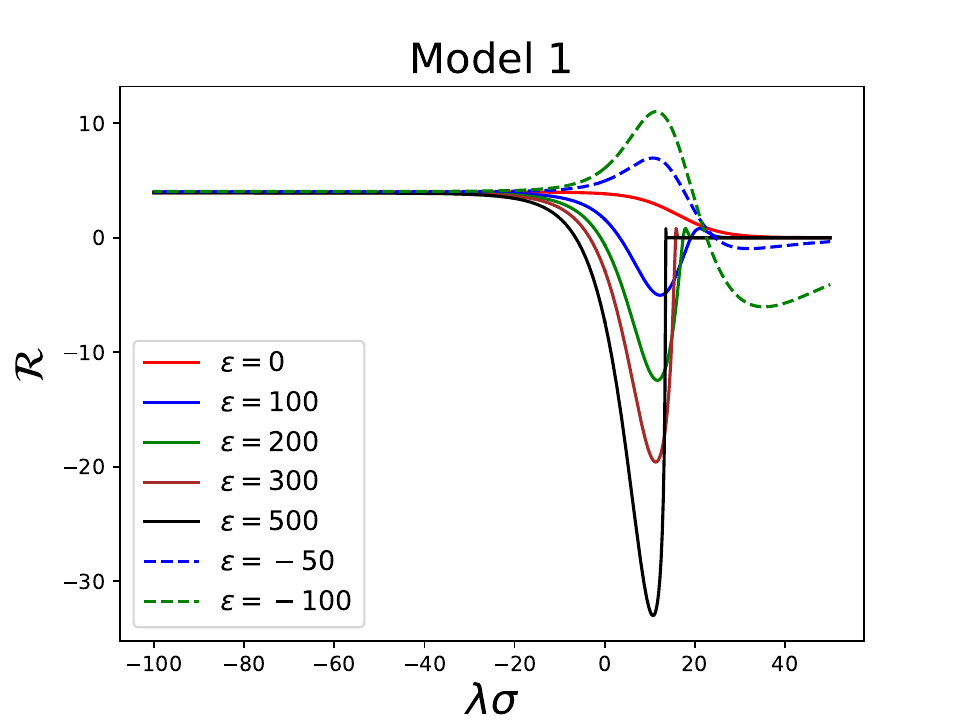}
  \end{subfigure}%
  \begin{subfigure}{.24\textwidth}
  \centering
    \includegraphics[width=1\linewidth]{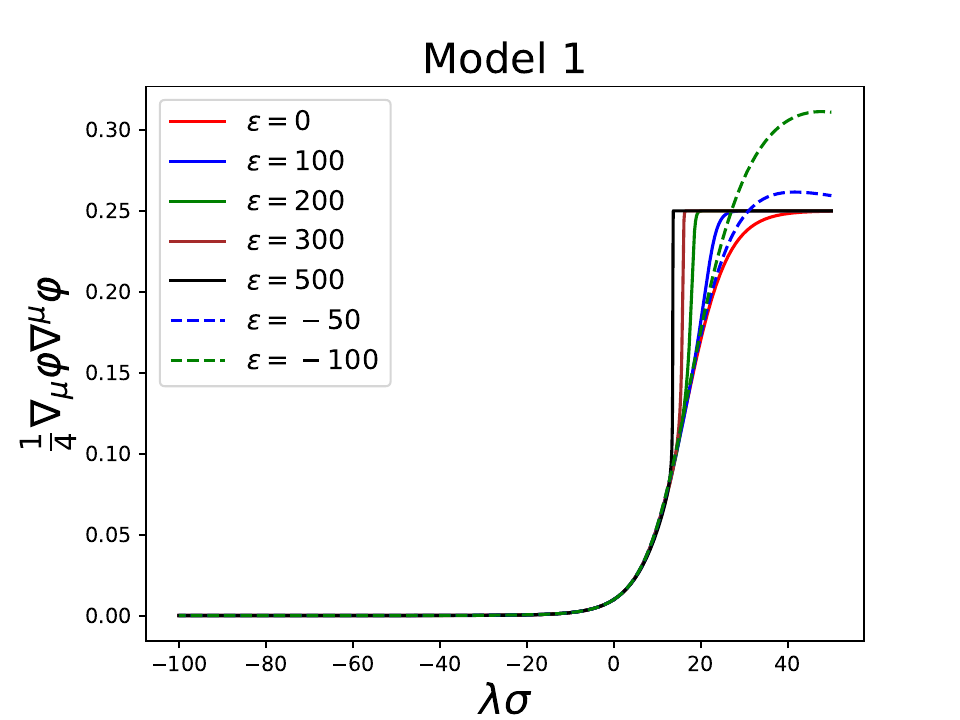}
  \end{subfigure}  
    \begin{subfigure}{.24\textwidth}
  \centering
    \includegraphics[width=1\linewidth]{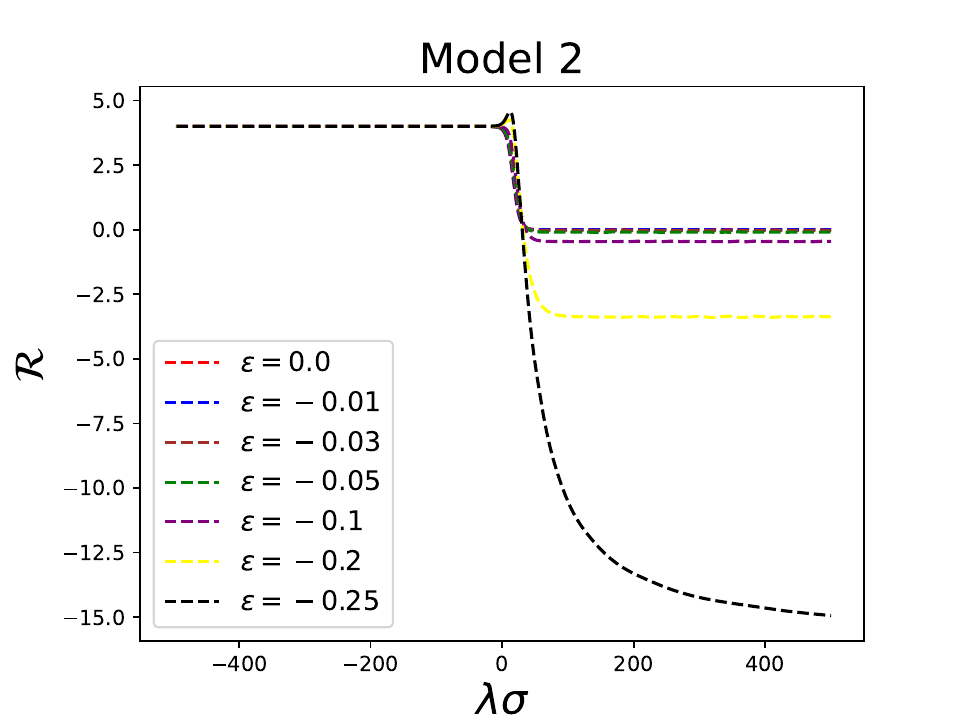}
  \end{subfigure}%
  \begin{subfigure}{.24\textwidth}
  \centering
    \includegraphics[width=1\linewidth]{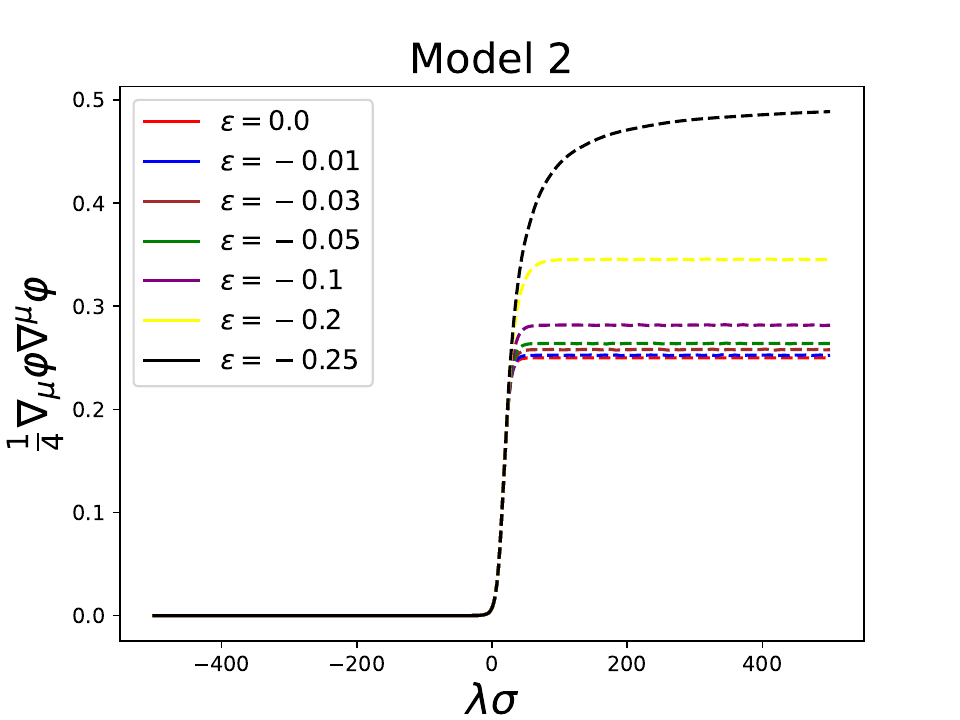}
  \end{subfigure}
  \caption{Numerical solution of the 2-D Ricci scalar and $X_{\varphi}$ for different values of $\epsilon$ for Model 1 and Model 2.}
\label{FIG.1}
\end{figure*}

It is challenging, if not impossible, to obtain exact and analytical solutions for these coupled non-linear differential equations. As we show below, the numerical solutions with high accuracy can provide crucial insights into the behavior of the classical solutions. In Fig. \eqref{FIG.1}, we plotted the solutions of field equations as a function of $\sigma$ for different values of $\epsilon$ for Models 1 and 2, respectively. Specifically, we have plotted 2-D Ricci scalar (${\cal R}$) and the kinetic term of the dilaton field ($-X_{\varphi}$). 

From Fig. \ref{FIG.1},  we infer the following salient features for Model 1: 
First, the 2-D Ricci scalar 
and $X_{\varphi}$ saturate at spatial infinity for various positive values of $\epsilon$ as $\sigma \to \infty$. 
Second, it is interesting to note that the saturated values do not change even when changing the values of the $\epsilon$ parameter. 
Third, to compare with the analytical expressions for CGHS, we have plotted these quantities for CGHS model $(\epsilon = 0)$. Our numerical results match with the analytical expression obtained in Ref.~\cite{potaux2023spacetime}:
\begin{equation}
\label{eq:CGHS-Ricciscalar}
\begin{split}
\nabla_{\mu}\varphi\nabla^{\mu}\varphi & = - \frac{\lambda^{4}x^{+}x^{-}}{M/\lambda - \lambda^{2}
x^{+}x^{-}} = \frac{\lambda^{2}e^{2\lambda\sigma}}{M/\lambda + e^{2\lambda\sigma}}\\
\mathcal{R} & = \frac{4M\lambda^{2}}{M - \lambda^{3}x^{+}x^{-}} = \frac{4M\lambda^{2}}{M + \lambda
e^{2\lambda\sigma}}
\end{split}
\end{equation} 
Thus, our numerical results provide correct results for all values of $\lambda \sigma$ and show that the corrections to CGHS do not lead to any divergences. 
Lastly, the asymptotic values for Model 1 can be understood from
the fact that for $\sigma\rightarrow\infty$, $\varphi$ becomes negative, making the CGHS action dominant over the correction terms. On the other hand, 
in $\sigma \rightarrow-\infty$, $\nabla_{\mu}\varphi\nabla^{\mu}\varphi$ saturates toward the zero value, which also means the corrections to the CGHS action become sub-dominant.

From Fig. \ref{FIG.1},  we infer the following salient features for Model 2: 
First, the numerical results match the analytical expressions \eqref{eq:CGHS-Ricciscalar} for $\epsilon = 0$. 
Second, unlike Model 1, in this model, the saturated values of $\mathcal{R}$ and $\frac{1}{4}\nabla_{\mu}\varphi\nabla^{\mu}\varphi$ change significantly at $\sigma\rightarrow\infty$ depending on the values of $\epsilon$. For the
numerical computation, we have considered negative values of $\epsilon$ to make the field theory stable, which follows from the action of this model. Further, $\epsilon > - 0.25$ to have a real-valued solution of $X$ as a function of $\varphi$. 

To understand the reason for the difference in the behavior of the various quantities in the two models, let us rewrite the equation of motion of the Horndeski action \eqref{eq:Horndeskiaction}  in the following form~\cite{1974-Horndeski-IJTP,Clifton:2011jh,Kobayashi:2019hrl}: 
\[
G_{\mu \nu} = 8 \pi G_{\rm eff} T_{\mu \nu}^{\rm Corr} \, ,
\]
where $T_{\mu \nu}^{\rm Corr}$ are the corrections to GR that are rewritten as matter corrections, and $G_{\rm eff}$ is the effective gravitational coupling. Due to the presence of a scalar field, $G_{\rm eff}$ may not be constant. Rewriting the Horndeski equations of motion in the above form leads~\cite{Zumalacarregui:2016pph}:
\begin{equation}
G_{\rm eff}^{-1} = 2\bigg(G_{4}-2XG_{4,X}+XG_{5,\Phi}\bigg),
\end{equation}
For the above three cases, we have:
\begin{eqnarray}
\text{Brans-Dicke} &~~~& G_{\rm eff}^{-1}=\Phi \\
\text{Corrected Model 1} &~~~& G_{\rm eff}^{-1}=\left(\Phi+X\mathcal{G}\right)  \\
\text{Corrected Model 2} &~~~& G_{\rm eff}^{-1}=\left(\Phi+\frac{X\mathcal{G}}{\Phi}\right)
\end{eqnarray}
The two models have different dependencies on $\Phi$, leading to different asymptotic values. Specifically, the corrected model-2 has a "duality" under the transformation $\Phi \to \frac{X\mathcal{G}}{\Phi}$. This leads to similar dependence for large $\Phi$ and small $\Phi$ (assuming that $X$ does not vanish). 

\noindent \section{Impact on Hawking radiation:} 
So far, our analysis of corrections to the CGHS model is purely classical and without any extra matter field. For the CGHS model, it is known that the BH formation is an inevitable outcome, regardless of the nature of the matter field. For instance, Callan etal~\cite{callan1992evanescent} considered a massless scalar field $f$
\begin{equation}\label{matter action}
S_{m} = - \frac{1}{4\pi}\int d^{2}x \, \sqrt{-g} \nabla_{\mu}f\nabla^{\mu}f.
\end{equation}  
and showed that a closed system of field equations can describe the evolution of dilaton gravity and scalar field. However, to our knowledge, a general analytical solution to the CGHS field equations with the scalar field is yet to be discovered. The numerical and approximate solutions have been obtained~\cite{Piran:1993tq,2010-Ori-PRD,Ramazanoglu:2010aj,Ashtekar:2010qz}. Given this, it may not be easy to obtain analytical solutions to the CGHS corrected models \eqref{eq:CGHScorrec}. However, it is possible to make certain concrete statements assuming that the black hole forms in these models in the presence of scalar field \eqref{matter action}.

Keeping this in mind, we qualitatively analyze the effect of higher derivative terms on the Hawking radiation for the 2-D metric~\eqref{eq:2Dmetric}. The conformal invariance of scalar fields causes the trace of the classical stress tensor to vanish; however, the quantum expectation value of the trace does not vanish. As a result, new source terms of quantum origin enter the geometry's equations of motion, changing the geometry's evolution. The new source terms lead to modifications in the evolution of the geometry~\cite{Fulling:1977jm,davies1977quantum,Robinson:2005pd,Banerjee:2008sn,1984-Birrell.Davies-Book,2009-Parker-Toms-Book}. Specifically, for the above scalar field action \eqref{matter action}, it has been shown that:
\begin{equation}
\langle T^{(f)} \rangle = \frac{1}{24 \pi} {\cal R};~~\langle T_{\pm\pm}^{(f)}\rangle = \frac{1}{12\pi}[\partial_{\pm}^{2}\omega - (\partial_{\pm}\omega)^{2} + t_{\pm}]
\end{equation} 
where $t_{\pm}$ can be obtained from the boundary conditions. For the CGHS corrected Model 1, the trace of the stress tensor is:
\begin{eqnarray}
\langle T^{(f)} \rangle  & = &  -\frac{1}{6\pi} 
\left(\Box\varphi - \nabla_{\mu}\varphi\nabla^{\mu}\varphi + \lambda^{2} \right) \\
 & +& \frac{1}{3 \pi} \Gamma G e^{2\varphi}[\Box\varphi\nabla_{\mu}\varphi\nabla^{\mu}\varphi + 2 \nabla^{\mu}
\varphi\nabla^{\nu}\varphi\nabla_{\mu}\nabla_{\nu}\varphi]. \nonumber
\end{eqnarray}
As mentioned earlier, obtaining the exact solutions for the CGHS model with matter is hard. Since the corrections are non-linear, obtaining the solutions is impossible, so the quantization is impractical. However, we can obtain the corrections to the CGHS by studying the linearized equations of motion \eqref{eq:EOM-CGHSCorrec}, leading to:
\begin{equation}
t_{\pm} -
t_{\pm}^{CGHS} =\epsilon[2\partial_{\pm}\bar{\omega}\partial_{\pm}\delta\omega - \partial_{\pm}^{2} \delta\omega] \simeq 2 \epsilon \partial_{\pm}\bar{\omega}\partial_{\pm}\delta\omega 
\label{eq:HawkingCorrec}
\end{equation}
where the initial conditions for $\bar{\omega}$ and $\delta\omega$ are set at $\sigma_+ \to - \infty$.  [For details, see Sec. VI in Ref.~\cite{2023-Mandal.etal-Supp}]. Note that $\partial^2 \delta\omega$ can be ignored in the linear regime. The corrections vanish if $\partial_{\pm} \delta \omega$ vanishes. In that case, the results confirm the earlier results of Jacobson and Unruh~\cite{jacobson1991black, unruh1995sonic} that the trans-Planckian signatures do not modify the spectrum. However, as discussed earlier, such a thing requires highly fine-tuned parameters ${\cal G}$ and $\lambda$. As mentioned, evaluation of $\langle T_{\pm\pm}^{f}\rangle$ exactly requires the exact solutions of the modified field equations in terms of null coordinates $\sigma^{\pm}$, which will be discussed in future work.

\noindent \section{Discussions:}
We have provided a systematic procedure to include a higher-derivative correction to the CGHS model and study the end stages of black hole evaporation. One of the advantages of this approach is that the higher-order corrections do not introduce Ostrogradsky instability (ghosts). Thus, the generic action \eqref{eq:CGHScorrec} can be used to understand the higher-derivative corrections to the s-wave of the 4-D Hawking radiation. Our analysis provides a first step to investigate the effects of Hawking radiation from PBH in the mass range $10^{16} - 10^{17}$~gm. Specifically, the positron annihilation rate implied by INTEGRAL's measurements of the Galactic 511 keV line 
has been used to place strong constraints on PBH in these mass ranges~\cite{DeRocco:2019fjq,Keith:2021guq}. However, these results assume that the Hawking flux for these ranges is the same as that of larger BHs. However, as discussed above \eqref{eq:HawkingCorrec}, there can be appreciable corrections to Hawking radiation for these mass ranges. Thus, our analysis can further constrain PBH as dark matter. 

Our analysis can possibly provide a way to understand the effect of higher-derivative terms on the page curve~\cite{Page:1993wv,Ho:2022gpg,Akhmedov:2023gqf}. For the BHs in GR, it is known that BH entropy increases monotonically until half its lifetime and decreases monotonically until the BH completely evaporates. If the higher-derivative corrections modify the Hawking flux for PBHs, there can likely be corrections to Page's analysis at the end stages of BH evaporation. This requires detailed numerical analysis along the lines of \cite{Ashtekar:2010qz,Ramazanoglu:2010aj}. This is currently under investigation. 

A mapping from 4d metric action (Einstein-Hilbert) to an effective two-dimensional model can be obtained through spherical reduction or via near horizon approximation~\cite{Louis-Martinez:1995aki,Carlip:2017xne,Carlip:2019dbu}. However, the two approaches are quite distinct. In the case of spherical reduction, the Einstein-Hilbert action reduces to a dilaton gravity theory considering the areal radius as a field. This reduction is valid every-where on the two-dimensional manifold and one can mimic the radial collapse of matter leading to stationary black hole. However, the near horizon approximation (of a non-degenerate horizon) describes the two-dimensional metric in the $t-r$ plane to be Rindler metric near the horizon. In this case, the areal radius do not play any role in determining the dynamics of the 2-D theory, and describes a stationary black hole solution. It will be interesting to explore the consequences of corrections on BMS symmetries~\cite{Carlip:2017xne,Carlip:2019dbu}.

For the quantum corrected 4-D reduced  Einstein-Hilbert action, Kazakov and Solodukhin showed that quantum corrections deform the line element of a Schwarzschild black hole leading to a non-singular space-time~\cite{Kazakov:1993ha}. It will be interesting to see whether the action \eqref{eq:CGHScorrec} can lead to non-singular space-time. The above analysis can be extended to other 2-D gravity models like Liouvelle gravity, JT gravity. These are currently under investigation. 

\emph{Acknowledgements:}
The authors thank Indranil Chakraborty, Saurya Das, S. Mahesh Chandran and Ashu Kushwaha for comments. The authors thank SERB grant (Project RD/0122-SERB000-044) for funding this research. The MHRD fellowship at IIT Bombay financially supports TP.

\bibliographystyle{apsrev}
\bibliography{Draft}

\pagebreak
\newpage

\textbf{Supplemental Material}

This document contains details related to the calculations presented in the main text

\appendix

\section{Spherical reduction of $d-$D Einstein-Hilbert Action}
\label{AppendixO}

In this Appendix we show that the spherical reduction of $(n + 2)-$D Einstein-Hilbert action
\begin{align}
\label{eq:dDEHaction-01}
 S_{\rm EH}^{(n + 2)} =\frac{1}{16\pi G_{n + 2}} \int d^{n + 2}x {\sqrt{-g^{(n + 2)}}}R^{\left(n + 2\right)}
\end{align}
can not lead to CGHS action (3). To go about this, we consider $(n + 2)-$dimensional manifold to be a product manifold of $\mathcal{M}^{\left(2\right)}\times S^{n}$,where, $\mathcal{M}^{\left(2\right)}$ is 
an arbitrary  2-D manifold and $S^{n}$ is a $n$-sphere. In other words, $(n + 2)-$ dimensional metric is given by
\begin{equation}
  ds^2=g_{AB} dx^{A}dx^{B}+\rho^2(x^A) d\Omega^2_{n}  \, ,
\end{equation}
Since, the metric is block diagonal and the 2-d is conformally flat, $(n + 2)-$D Ricci scalar can be written as:
\begin{equation}
R^{(n + 2)} = {\cal R} + \frac{n(3-n)}{\rho^2}(\nabla\rho)^2-\frac{n}{\rho^2}\Box\rho^2+\frac{n(n-1)}{\rho^2}  \, .
\label{eq:dDEHaction-02}
\end{equation}
Substituting the above form in Eq.~\eqref{eq:dDEHaction-01}, we have:
\begin{align}
S &= \frac{\Omega_{n}}{16\pi G_{n + 2}} \int dx^A \sqrt{-g}\rho^n 
\left[ {\cal R} + \frac{n(3-n)}{\rho^2}(\nabla\rho)^2 \right.  \\
& \left. -\frac{n}{\rho^2}\Box\rho^2 +\frac{n(n-1)}{\rho^2} \right] \nonumber
\label{eq:dDEHaction-03}
\end{align}
where $\Omega_n$ is the area of the $n-$Sphere. To map to CGHS action (3), setting 
 \begin{equation}
 G_{n + 2}=\frac{\Omega_n}{8\lambda^n} ; ~~\rho=\frac{e^{-\frac{2}{n}\phi}}{\lambda}  
 \end{equation}
in the above action, we get:
 \begin{equation}
 S=\int \frac{d^2x}{2\pi} \sqrt{-g} e^{-2\phi}\left[ {\cal R} + 
 \frac{4(3-n)}{n}(\nabla\phi)^2+n(n-1)\lambda^2 e^{\frac{4}{n}\phi}\right]  \nonumber
 \end{equation}
The second term in the above action matches with CGHS action (3) 
only when $n = 3/2$. Since, $n$ has to be an integer, we infer that the dimensional reduction of  Einstein-Hilbert action in arbitrary dimensions can not lead to the CGHS action.

Moreover, we can start with the Einstein-Hilbert action in $n$ dimensions, given by
\begin{equation}
    S=\frac{1}{16\pi G_{n+2}}\int d^{n+2}x\sqrt{-g^{(n+2)}}R^{(n+2)}.
\end{equation}
The spherical symmetric ansatz for the line-element is considered to be the following:
\begin{equation}
    ds^2=g_{AB}(x^{C})dx^{A}dx^{B}+\rho^{2}(x^{A})d\Omega^2_{n}.
\end{equation}
After performing the angular integration, we obtain
\begin{equation}
\label{3}
    S=\frac{\Omega_{n}}{16\pi G_{n+2}}\int d^2x \sqrt{-g^{(2)}}\rho^{n}\left[{\cal R}^{(2)}+\frac{n(3-n)}{\rho^2}(\nabla\rho)^2-\frac{n}{\rho^2}\Box{\rho^2}+\frac{n(n-1)}{\rho^2}\right].
\end{equation}
Now, we can perform a Weyl transformation on the $2D$ metric, which is described by
\begin{equation}
    g_{AB}=\frac{1}{f^{2}\left(\frac{\rho}{\alpha}\right)}\Tilde{g}_{AB},
\end{equation}
where a dimensionful constant $\alpha$ is introduced such that $f(\rho/\alpha)$ is dimensionless.
The Ricci scalars are then related as
\begin{equation}
    {\cal R}^{(2)} = f^{2}\left(\frac{\rho}{\alpha}\right)\Tilde{{\cal R}}^{2}+2\Tilde{g}^{AB}\Tilde{\nabla}_{A}\left(f\left(\frac{\rho}{\alpha}\right)\Tilde{\nabla}_{B}f\left(\frac{\rho}{\alpha}\right)\right)-4\left(f^{\prime}\left(\frac{\rho}{\alpha}\right)\right)^2(\Tilde{\nabla}\rho)^2.
\end{equation}
Using the above relation, equation (\ref{3}) becomes the following
\begin{eqnarray}
    S &=& \frac{\Omega_{n}}{16\pi G_{n+2}}\int d^2x \sqrt{-\Tilde{g}^{(2)}}\frac{1}{f^{2}\left(\frac{\rho}{\alpha}\right)}\rho^{n}\bigg[f^{2}\left(\frac{\rho}{\alpha}\right)\Tilde{{\cal R}}^{(2)}+2\Tilde{g}^{AB}\Tilde{\nabla}_{A}\left(f\left(\frac{\rho}{\alpha}\right)\Tilde{\nabla}_{B}f\left(\frac{\rho}{\alpha}\right)\right)\nonumber \\ &&  -4(f^{\prime}\left(\frac{\rho}{\alpha}\right))^2(\Tilde{\nabla}\rho)^2+\frac{n(3-n)}{\rho^2}(\Tilde{\nabla}\rho)^2-\frac{n}{\rho^2}f^{2}\left(\frac{\rho}{\alpha}\right)\tilde{\Box}\rho^2+\frac{n(n-1)}{\rho^2}\bigg].
\end{eqnarray}
To get rid of the $\Tilde{\Box}\rho^2$ term, we choose $n=2$, and to get rid of the second term in the bracket as a surface term, we choose 
\begin{equation}
    f\left(\frac{\rho}{\alpha}\right)=\frac{\rho}{\alpha}.
\end{equation}
With the above conditions, we obtain the following form of the action:
\begin{equation}
    S=\frac{1}{4G}\int d^2x \sqrt{-\Tilde{g}^{(2)}}\bigg[\rho^2\Tilde{{\cal R}}^{(2)}-4(\Tilde{\nabla}\rho)^2+\frac{2\alpha^2}{\rho^2}(\Tilde{\nabla}\rho)^2+\frac{2\alpha^2}{\rho^2}\bigg].
\end{equation}
With the following substitution,
\begin{equation}
   \rho=\frac{e^{-\phi}}{\lambda}, 
\end{equation}
we obtain
\begin{equation}
    S=\frac{1}{4G}\int d^2x\sqrt{-g^{(2)}}\bigg[\frac{e^{-2\phi}}{\lambda^2}\Tilde{{\cal R}}^{(2)}-4\frac{e^{-2\phi}}{\lambda^2}(\Tilde{\nabla}\phi)^2+2\alpha^2(\Tilde{\nabla}\phi)^2+2\alpha^2\lambda^2e^{2\phi}\bigg].
\end{equation}
From the above action, we can infer that this is not the same as the CGHS model. Hence, the conformal transformation in the $\{x^{A}\}$ 2-D space-time will not lead to the CGHS action.

\section{Beyond Classical CGHS action: Details}
\label{AppendixOO}

We start with the Horndeski action (4) and introduce the the following quantities
\begin{equation}
\begin{split} 
& G_{2}(X, \Phi) = \Omega(\Phi)X - V(\Phi);~
 G_{3}(X, \Phi) = Xf(\Phi),\\  
 & G_{4}(\Phi) = \Phi, \Omega(\Phi) = - \frac{6}{\Phi} + \frac{h(\Phi)}{\Lambda^{2}}.   
\end{split}    
\end{equation}
With the above-mentioned choice, we obtain the following expression of the action
\begin{equation}
\begin{split}
S & = \frac{1}{16\pi G}\int\sqrt{-g^{(4)}}d^{4}x\Big[ - \frac{6}{\Phi}X - V(\Phi) + \Phi R^{(4)} + \frac{h(\Phi)}{\Lambda^{2}}X\\
 & - f(\Phi)X\Box\Phi - G_{5}(\Phi)G_{\mu\nu}^{(4)}\nabla^{\mu}\Phi\nabla^{\nu}\Phi\Big]
 \end{split}    
\end{equation}
We can decomposed the above action as $S = S_{0} + S_{1}$ where
\begin{equation}
\begin{split}
S_{0} & = \frac{1}{16\pi G}\int\sqrt{-g^{(4)}}d^{4}x \Big[ - \frac{6}{\Phi}X - V(\Phi) + \Phi R^{(4)}\Big]\\
S_{1} & = \frac{1}{16\pi G}\int\sqrt{-g^{(4)}}d^{4}x \Big[\frac{h(\Phi)}{\Lambda^{2}}X - f(\Phi)X\Box\Phi  \\ 
& - G_{5}(\Phi)G_{\mu\nu}^{(4)}\nabla^{\mu}\Phi\nabla^{\nu}\Phi\Big].
\end{split}
\end{equation}
Considering the following constraints
\begin{equation}
\rho = \Lambda e^{- 2\varphi}, \ \Phi = e^{2\varphi},    
\end{equation}
we obtain the following relations
\begin{equation}
\begin{split}
X & = - \frac{1}{2}\nabla_{\mu}\Phi\nabla^{\mu}\Phi = - 2e^{4\varphi}\nabla_{\mu}\varphi\nabla^{\mu}\varphi = 4e^{4\varphi}X_{\varphi}\\
6\frac{X}{\Phi} & = 24 e^{2\varphi}X_{\varphi},
\end{split}    
\end{equation}
and we choose the field potential $V(\Phi)$ to be of the following form
\begin{equation}
V(\Phi) = - 4\lambda^{2}\Phi + \frac{2}{\Lambda^{2}}\Phi^{3}.    
\end{equation}
As a result, the action $S_{0}$ reduces to the following form
\begin{equation}
\begin{split}
S_{0} & = \frac{4\pi}{16\pi G}\int\sqrt{-g}d^{2}x \rho^{2}\Big[ - 24 e^{2\varphi} X_{\varphi} + 4\lambda^{2} e^{2\varphi} - \frac{2}{\Lambda^{2}} e^{6\varphi}\\
 & + \Phi\left(\mathcal{R} + \frac{2}{\rho^{2}} + \frac{2}{\rho^{2}}\nabla_{\mu}\rho\nabla^{\mu}\rho - \frac{2}{\rho^{2}}\Box\rho^{2}\right)\Big]
\end{split}
\end{equation}
where
\begin{equation}
\begin{split}
\frac{2}{\rho^{2}} & = \frac{2}{\Lambda^{2}}e^{4\varphi}, \ \frac{2}{\rho^{2}}\nabla_{\mu}\rho\nabla^{\mu}\rho = - 16X_{\varphi}\\
- \frac{2}{\rho^{2}} & \Box\rho^{2} = - \frac{2}{\rho^{2}}(2\nabla_{\mu}\rho\nabla^{\mu}\rho + 2\rho\Box\rho)\\
 & = - 16\nabla_{\mu}\varphi\nabla^{\mu}\varphi - \frac{4}{\rho}\Box\rho\\
 & = 32X_{\varphi} - \frac{4}{\Lambda} e^{2\varphi}\Box\rho.
\end{split}    
\end{equation}
Hence, the action reduces to the following form
\begin{equation}
\begin{split}
S_{0} & = \frac{\Lambda^{2}}{4G}\int\sqrt{-g}d^{2}x \ e^{-4\varphi} \Big[- 24 e^{2\varphi}X_{\varphi} + 4\lambda^{2}e^{2\varphi} - \frac{2}{\Lambda^{2}}e^{6\varphi}\\
 & + e^{2\varphi}\left(\mathcal{R} + \frac{2}{\Lambda^{2}}e^{4\varphi} - 16X_{\varphi} + 32X_{\varphi} - 4 e^{2\varphi}\Box\rho\right)\Big]
\end{split}
\end{equation}
in which we neglect the last term as it gives rise to a surface term. Therefore, we obtain the following expression
\begin{equation}
S_{0} = \frac{\Lambda^{2}}{4G}\int\sqrt{-g} d^{2}x e^{-2\varphi}[\mathcal{R} - 8X_{\varphi} + 4\lambda^{2}] = \frac{\Lambda^{2}\pi}{G}S_{CGHS}.    
\end{equation}
On the other hand, the $S_{1}$ part of the action can be expressed as
\begin{equation}\label{CGHS correction}
\begin{split}
S_{1} 
 & = \frac{1}{4G}\int\sqrt{-g}d^{2}x \rho^{2}\Big[\frac{h(\Phi)}{\Lambda^{2}}X - f(\Phi)X\Box\Phi \\
 &- G_{5}(\Phi)G_{\mu\nu}^{(4)}\nabla^{\mu}\Phi\nabla^{\nu}\Phi\Big]
\end{split}    
\end{equation}
In the above expression, the last term can be expressed
\begin{equation}
\begin{split}
- \rho^{2} & G_{5}(\Phi)G_{\mu\nu}^{(4)}\nabla^{\mu}\Phi\nabla^{\nu}\Phi\\
 & = - \Lambda^{2}e^{-4\varphi}G_{5}(\Phi)\left(R_{\mu\nu}^{(4)} - \frac{1}{2}g_{\mu\nu}^{(4)}R^{(4)}\right)\nabla^{\mu}\Phi\nabla^{\nu}\Phi\\
 & = - \Lambda^{2}G_{5}(\Phi)\Big[16\nabla_{\mu}\nabla_{\nu}\varphi\nabla^{\mu}\varphi\nabla^{\nu}\varphi + \frac{8}{\Lambda^{2}}e^{4\varphi}X_{\varphi} \\ 
 & + 64X_{\varphi}^{2} + 32\Box\varphi X_{\varphi}\Big].
\end{split}
\end{equation}
On the other hand, the first term in (\ref{CGHS correction}) can be expressed as
\begin{equation}
\begin{split}
\rho^{2}\frac{h(\Phi)}{\Lambda^{2}}X & = - \frac{1}{2}\nabla_{\mu}\Phi\nabla^{\mu}\Phi \times e^{-4\varphi}h(\Phi)\\
 & = - 2\nabla_{\mu}\varphi\nabla^{\mu}\varphi h(\Phi)\\
 & = 4h(\Phi)X_{\varphi}
\end{split}    
\end{equation}
whereas the second term can be expressed as
\begin{equation}
\begin{split}
-\rho^{2}f(\Phi)X\Box\Phi & = \frac{1}{2}\rho^{2}f(\Phi)\nabla_{\mu}\Phi\nabla^{\mu}\Phi \Box\Phi\\
 & = - 8\Lambda^{2}e^{2\varphi}f(\Phi)X_{\varphi}[\Box\varphi - 4X_{\varphi}]
\end{split}
\end{equation}
Therefore, the correction to the CGHS action reduces to the following form
{\small 
\begin{equation}
\begin{split}
S_{1} & = \frac{\Lambda^{2}}{4G}\int\sqrt{-g}d^{2}x \Big[\frac{4h(\Phi)}{\Lambda^{2}}X_{\varphi} - 8e^{2\varphi}f(\Phi)X_{\varphi}[\Box\varphi - 4X_{\varphi}]\\
 & - G_{5}(\Phi)\left(16\nabla_{\mu}\nabla_{\nu}\varphi\nabla^{\mu}\varphi\nabla^{\nu}\varphi + \frac{8}{\Lambda^{2}}e^{4\varphi}X_{\varphi} \right. \\
& \left.  + 64X_{\varphi}^{2} + 32\Box\varphi X_{\varphi}\right)\Big]    
\end{split}    
\label{AppB:eq02}
\end{equation}
}
{\bf For Model 1}, we choose
\begin{equation}
h(\Phi) = 2\mathcal{G}\Phi^{2}, \ f(\Phi) = - \frac{6\mathcal{G}}{\Phi}, \ G_{5} = \mathcal{G} \, .    
\end{equation}
Using the following relation, we have:
{\small
\begin{equation}
\begin{split}
& \int\sqrt{-g}d^{2}x  \nabla_{\mu}\nabla_{\nu}\varphi\nabla^{\mu}\varphi\nabla^{\nu}
\varphi = \\
& \int\sqrt{-g}d^{2}x \Big[\nabla_{\mu}(\nabla_{\nu}\varphi\nabla^{\nu}\varphi\nabla^{\mu}\varphi) - \nabla_{\nu}\varphi\nabla^{\nu}\varphi\Box\varphi \\ 
& - \nabla_{\nu}\varphi\nabla^{\mu}\varphi\nabla_{\mu}\nabla^{\nu}\varphi\Big]\\
& \implies\int\sqrt{-g}d^{2}x  \nabla_{\mu}\nabla_{\nu}\varphi\nabla^{\mu}\varphi\nabla^{\nu}
\varphi = \int\sqrt{-g}d^{2}x X_{\varphi}\Box\varphi 
\end{split}
\end{equation}
}
Hence, the action reduces to the following form
\begin{equation}
\begin{split}
S_{1} & = \frac{\Lambda^{2}}{4G}\int\sqrt{-g}d^{2}x \Big[\frac{4h(\Phi)}{\Lambda^{2}}X_{\varphi} - 8e^{2\varphi}f(\Phi)X_{\varphi}(\Box\varphi - 4X_{\varphi})\\
 & - G_{5}\left(48X_{\varphi}\Box\varphi + \frac{8}{\Lambda^{2}}e^{4\varphi}X_{\varphi} + 64X_{\varphi}^{2}\right)\Big]\\
 & = \frac{\Lambda^{2}}{4G}\int\sqrt{-g}d^{2}x \Big[\frac{8\Phi^{2}\mathcal{G}}{\Lambda^{2}}X_{\varphi} + 48\mathcal{G}X_{\varphi}(\Box\varphi - 4X_{\varphi})\\
 & - \mathcal{G}\left(48X_{\varphi}\Box\varphi + \frac{8}{\Lambda^{2}}e^{4\varphi}X_{\varphi} + 64X_{\varphi}^{2}\right)\Big]\\
 & = - \frac{\Lambda^{2}}{4G}\int\sqrt{-g}d^{2}x \ 256\mathcal{G}X_{\varphi}^{2}
\end{split}    
\end{equation}
{\bf In Model 2}, we consider the following 
\begin{equation}
f(\Phi) = - \frac{6\mathcal{G}}{\Phi^{2}}, \ h(\Phi) = 2\mathcal{G}\Phi, \ G_{5}(\Phi) = \frac{\mathcal{G}}{\Phi}    
\end{equation}
which leads to the following expression of $S_{1}$
{\small 
\begin{equation}
\begin{split}
S_{1} & = \frac{\Lambda^{2}}{4G}\int\sqrt{-g}d^{2}x \ \Big[\frac{4h(\Phi)}{\Lambda^{2}}X_{\varphi} - 8e^{2\varphi}f(\Phi)X_{\varphi}(\Box\varphi - 4X_{\varphi})\\
 & - G_{5}(\Phi)\left(16\nabla_{\mu}\nabla_{\nu}\varphi\nabla^{\mu}\varphi\nabla^{\nu}\varphi + \frac{8}{\Lambda^{2}}e^{4\varphi}X_{\varphi} + 64X_{\varphi}^{2} + 32\Box\varphi X_{\varphi}\right)\Big]\\
 & = \frac{\Lambda^{2}}{4G}\int\sqrt{-g}d^{2}x \ \Big[\frac{8\mathcal{G}}{\Lambda^{2}}e^{2\varphi}X_{\varphi} + 48e^{-2\varphi}\mathcal{G}X_{\varphi}(\Box\varphi - 4X_{\varphi})\\
 & - \mathcal{G}e^{-2\varphi}\left(16\nabla_{\mu}\nabla_{\nu}\varphi\nabla^{\mu}\varphi\nabla^{\nu}\varphi + \frac{8}{\Lambda^{2}}e^{4\varphi}X_{\varphi} + 64X_{\varphi}^{2} + 32\Box\varphi X_{\varphi}\right)\Big]\\
 & = \frac{\Lambda^{2}}{4G}\int\sqrt{-g}d^{2}x \Big[48e^{-2\varphi}\mathcal{G}X_{\varphi}(\Box\varphi - 4X_{\varphi}) \\ 
& - \mathcal{G}e^{-2\varphi}\left(16\nabla_{\mu}\nabla_{\nu}\varphi\nabla^{\mu}\varphi\nabla^{\nu}\varphi + 64X_{\varphi}^{2} + 32\Box\varphi X_{\varphi}\right)\Big]
\end{split}    
\end{equation}
}
Using the following relation
{\small
\begin{equation}
\begin{split}
& \int\sqrt{-g}d^{2}x \ e^{-2\varphi} \nabla_{\mu}\nabla_{\nu}\varphi\nabla^{\mu}\varphi\nabla^{\nu}\varphi \\ 
& = \int\sqrt{-g}d^{2}x \ e^{-2\varphi}\left(- \nabla_{\mu}\varphi\nabla^{\mu}\varphi\Box\varphi - \nabla_{\mu}\nabla_{\nu}\varphi\nabla^{\mu}\varphi\nabla^{\nu}\varphi + 2(\nabla_{\mu}\varphi\nabla^{\mu}\varphi)^{2}\right)\\
& \implies \int\sqrt{-g}d^{2}x  \ e^{-2\varphi} \nabla_{\mu}\nabla_{\nu}\varphi\nabla^{\mu}\varphi\nabla^{\nu}\varphi \\ 
& = \int\sqrt{-g}d^{2}x \ e^{-2\varphi}\left(- \frac{1}{2}\nabla_{\mu}\varphi\nabla^{\mu}\varphi\Box\varphi + (\nabla_{\mu}\varphi\nabla^{\mu}\varphi)^{2}\right)\\
 & = \int\sqrt{-g}d^{2}x \ e^{-2\varphi} [X_{\varphi}\Box\varphi + 4X_{\varphi}^{2}]
\end{split}
\end{equation}
}
As a result, we obtain the following expression for $S_{1}$
\begin{equation}
\begin{split}
S_{1} & = \frac{\Lambda^{2}}{4G}\int\sqrt{-g}d^{2}x \ e^{-2\varphi}\Big[48\mathrm{G}X_{\varphi}(\Box\varphi - 4X_{\varphi}) \\ 
& - \mathcal{G}[48X_{\varphi}\Box\varphi + 128X_{\varphi}^{2}]\Big]\\
 & = - \frac{\Lambda^{2}}{4G}\int\sqrt{-g}d^{2}x \ e^{-2\varphi}320\mathcal{G}X_{\varphi}^{2}.
\end{split}   
\end{equation}
%


%

\section{Covariant field equations of derivative corrected CGHS action}\label{Appendix A}

In this appendix, we explicitly derive the covariant field equations of the corrected CGHS action \eqref{AppB:eq02} for the two models. Note that the constant $\mathcal{G}$ is a dimensionful quantity of length dimension $2$ and therefore, we may write the constant $\mathcal{G} = 16\pi G\mathcal{G}_{0}$
where $\mathcal{G}_{0}$ is a dimensionless constant. 

\subsection{Model 1}

In this section, we show the variation of the total action (CGHS + Horndeski) 
\begin{equation}
\begin{split}
S_{tot} & = \Lambda^{2}(\bar{S}_{0} + \bar{S}_{\text{corr}})\\ 
\bar{S}_{0} & = \frac{1}{4G}\int\sqrt{-g}d^{2}x e^{-2\varphi}\Big[\mathcal{R} + 4\nabla_{\mu}
\varphi\nabla^{\mu}\varphi + 4\lambda^{2}\Big]\\
\bar{S}_{\text{corr}} & = - \Gamma\int\sqrt{-g}d^{2}x (\nabla_{\mu}\varphi\nabla^{\mu}\varphi)
^{2}, \ \Gamma = 256\pi\mathcal{G}_{0},
\end{split}
\end{equation}
\textit{w.r.t} both the metric and the dilaton field $\varphi$. We obtain the following 
results for the variation of action \textit{w.r.t} the dilaton field
\begin{equation}
\begin{split}
\frac{\delta\bar{S}_{0}}{\delta\varphi} & = - \sqrt{-g^{(2)}}\frac{e^{-2\varphi}}{2G}[\mathcal{R}
 + 4\Box\varphi - 4\nabla_{\mu}\varphi\nabla^{\mu}\varphi + 4\lambda^{2}]\\
\frac{\delta\bar{S}_{\text{corr}}}{\delta\varphi} & = 4\Gamma\sqrt{-g^{(2)}}[\Box\varphi\nabla_{\mu}
\varphi\nabla^{\mu}\varphi + 2\nabla^{\mu}\varphi\nabla^{\nu}\varphi\nabla_{\mu}\nabla_{\nu}
\varphi],
\end{split}
\end{equation}
which implies that we obtain the following field equation from the variation of the total 
action \textit{w.r.t} the dilaton field 
\begin{equation}\label{dilaton varying field eqn}
\begin{split}
\mathcal{R} & + 4\Box\varphi - 4\nabla_{\mu}\varphi\nabla^{\mu}\varphi + 4\lambda^{2}\\
 & = 8\Gamma G e^{2\varphi}[\Box\varphi\nabla_{\mu}\varphi\nabla^{\mu}\varphi + 2\nabla^{\mu}
\varphi\nabla^{\nu}\varphi\nabla_{\mu}\nabla_{\nu}\varphi].
\end{split}
\end{equation}
On the other hand, we obtain the following results upon varying the action \textit{w.r.t} 
the metric
\begin{equation}
\begin{split}
\frac{2}{\sqrt{-g^{(2)}}}\frac{\delta\bar{S}_{0}}{\delta g^{\mu\nu}} & = \frac{e^{-2\varphi}}{G}
[\nabla_{\mu}\nabla_{\nu}\varphi + g_{\mu\nu}^{(2)}\{\nabla_{\alpha}\varphi\nabla^{\alpha}\varphi
\\
 & - \Box\varphi - \lambda^{2}\}]\\
\frac{2}{\sqrt{-g^{(2)}}}\frac{\delta\bar{S}_{\text{corr}}}{\delta g^{\mu\nu}} & = - \Gamma
[4\nabla_{\mu}\varphi\nabla_{\nu}\varphi\nabla_{\alpha}\varphi\nabla^{\alpha}\varphi \\
& - g_{\mu\nu}^{(2)}(\nabla_{\alpha}\varphi\nabla^{\alpha}\varphi)^{2}],
\end{split}
\end{equation}
which leads to the following field equation
\begin{equation}\label{metric varying field eqn}
\begin{split}
& \nabla_{\mu}\nabla_{\nu}\varphi  + g_{\mu\nu}^{(2)}(\nabla_{\alpha}\varphi\nabla^{\alpha}\varphi
 - \Box\varphi - \lambda^{2})\\
 & = G\Gamma e^{2\varphi}[4\nabla_{\mu}\varphi\nabla_{\nu}\varphi\nabla_{\alpha}\varphi
 \nabla^{\alpha}\varphi - g_{\mu\nu}^{(2)}(\nabla_{\alpha}\varphi\nabla^{\alpha}\varphi)^{2}]
\end{split}
\end{equation}

\subsection{Model 2}

In this section, we consider the Model 2 class of corrections to the CGHS model which is described
by the following action
\begin{equation}
\begin{split}
\mathcal{A}_{tot} & = \Lambda^{2}(\mathcal{A}_{0} + \mathcal{A}_{\text{corr}})\\
\mathcal{A}_{0} & = \frac{1}{4G}\int\sqrt{-g}d^{2}x e^{-2\varphi}\Big[\mathcal{R} + 4\nabla_{\mu}
\varphi\nabla^{\mu}\varphi + 4\lambda^{2}\Big]\\
\mathcal{A}_{\text{corr}} & = - \bar{\Gamma}\int\sqrt{-g}d^{2}x e^{-2\varphi}(\nabla_{\mu}\varphi
\nabla^{\mu}\varphi)^{2}, \ \bar{\Gamma} = 320\pi\mathcal{G}_{0}.
\end{split}
\end{equation}
Varying the above action \textit{w.r.t} the dilaton field $\varphi$, we obtain the following
equation of motion
\begin{equation}
\begin{split}
0 & = \frac{1}{\Lambda^{2}}\frac{\delta\mathcal{A}_{tot}}{\delta\varphi}\frac{1}{\sqrt{-g}} = 
- \frac{e^{-2\varphi}}{G}(\mathcal{R} + 4\nabla_{\mu}\varphi\nabla^{\mu}\varphi + 4\lambda^{2})\\
 & + 2\bar{\Gamma}(\nabla_{\mu}\varphi\nabla^{\mu}\varphi)^{2}e^{-2\varphi} - \frac{2}{G}\nabla_{\mu}
 (e^{-2\varphi}\nabla^{\mu}\varphi)\\
 & + 4\bar{\Gamma}\nabla_{\mu}(e^{-2\varphi}\nabla^{\mu}\varphi\nabla_{\nu}\varphi\nabla^{\nu}\varphi),
\end{split}
\end{equation}
which can be re-expressed as follows
\begin{equation}
\begin{split}
& \mathcal{R}  + 4\Box\varphi - 4\nabla_{\mu}\varphi\nabla^{\mu}\varphi + 4\lambda^{2} = - 12
\bar{\Gamma}G(\nabla_{\mu}\varphi\nabla^{\mu}\varphi)^{2}\\
 & + 8\bar{\Gamma}G\Box\varphi\nabla_{\nu}\varphi\nabla^{\nu}\varphi + 16\bar{\Gamma}G\nabla_{\mu}
 \varphi\nabla_{\nu}\varphi\nabla^{\mu}\nabla^{\nu}\varphi. 
\end{split}
\end{equation}
On the other hand, varying the action \textit{w.r.t} the matter field, we obtain the following
relations
\begin{equation}
\begin{split}
\frac{2}{\sqrt{-g}} & \frac{\delta\mathcal{A}_{0}}{\delta g^{\mu\nu}} = 
\frac{e^{-2\varphi}}{G}[\nabla_{\mu}\nabla_{\nu}\varphi + g_{\mu\nu}(\nabla_{\alpha}\varphi
\nabla^{\alpha}\varphi - \Box\varphi - \lambda^{2})]\\
\frac{2}{\sqrt{-g}} & \frac{\delta\mathcal{A}_{\text{corr}}}{\delta g^{\mu\nu}} = - \bar{\Gamma}
e^{-2\varphi}[4\nabla_{\mu}\varphi\nabla_{\nu}\varphi\nabla_{\alpha}\varphi\nabla^{\alpha}\varphi \\
& - g_{\mu\nu}(\nabla_{\alpha}\varphi\nabla^{\alpha}\varphi)^{2}],
\end{split}
\end{equation}
and combining the following two relations, we obtain the following field equations
\begin{equation}
\begin{split}
\nabla_{\mu}\nabla_{\nu}\varphi & + g_{\mu\nu}(\nabla_{\alpha}\varphi\nabla^{\alpha}\varphi - 
\Box\varphi - \lambda^{2})\\
= \bar{\Gamma}G & [4\nabla_{\mu}\varphi\nabla_{\nu}\varphi\nabla_{\alpha}\varphi\nabla^{\alpha}
\varphi - g_{\mu\nu}(\nabla_{\alpha}\varphi\nabla^{\alpha}\varphi)^{2}].
\end{split}
\end{equation}

\section{Field equations in null coordinates}\label{Appendix B}

In this appendix, we obtain the equations of motion for both the models in-terms of null coordinates. Specifically, we choose a gauge in which the 2-D line-element can be expressed as
\begin{equation}
ds^{2} = - e^{2\omega}dx^{+}dx^{-}
\end{equation}
which implies
\begin{equation}
\begin{split}
g_{+ -} & = g_{- +} = - \frac{1}{2}e^{2\omega}, \ g^{+ -} = g^{- +} = - 2e^{-2\omega}\\
g_{++} & = g_{--} = g^{++} = g^{--} = 0.
\end{split} 
\end{equation}
The only non-zero Christoffel symbols associated with this line-element are the following
\begin{equation}
\Gamma_{ \ ++}^{+} = 2\partial_{+}\omega, \ \Gamma_{ \ --}^{-} = 2\partial_{-}\omega.
\end{equation}
With the help of the above-mentioned null coordinates, we obtain the following results
\begin{equation}\label{useful relations}
\begin{split}
\mathcal{R} & = 8e^{-2\omega}\partial_{+}\partial_{-}\omega\\
\Box\varphi & = - 4e^{-2\omega}\partial_{+}\partial_{-}\varphi\\
\nabla_{\mu}\varphi\nabla^{\mu}\varphi & = - 4e^{-2\omega}\partial_{+}\varphi\partial_{-}
\varphi\\
\nabla_{\mu}\nabla_{\nu}\varphi\nabla^{\mu}\varphi\nabla^{\nu}\varphi & = 4e^{-4\omega}
[\partial_{+}^{2}\varphi(\partial_{-}\varphi)^{2} - 2\partial_{+}\omega\partial_{+}(\partial_{-}
\varphi)^{2}\\
 & + 2\partial_{+}\partial_{-}\varphi\partial_{+}\varphi\partial_{-}\varphi + 2\partial_{-}^{2}
 \varphi(\partial_{+}\varphi)^{2}\\
 & - 2\partial_{-}\omega\partial_{-}\varphi(\partial_{+}\varphi)^{2}]\\
\nabla_{\mu}\varphi\nabla^{\mu}\varphi\Box\varphi & = 16e^{-4\omega}\partial_{+}\partial_{-}
\varphi\partial_{+}\varphi\partial_{-}\varphi. 
\end{split}
\end{equation}

\subsection{Model 1}

Using the above expressions, we obtain the following constraint equations which are the $++$
and $--$ component of metric varying field equations
\begin{equation}\label{constraint eqns}
\begin{split}
\partial_{+}^{2}\varphi & = 2\partial_{+}\omega\partial_{+}\varphi - 16G\Gamma 
e^{2(\varphi - \omega)}(\partial_{+}\varphi)^{3}\partial_{-}\varphi\\
\partial_{-}^{2}\varphi & = 2\partial_{-}\omega\partial_{-}\varphi - 16G\Gamma 
e^{2(\varphi - \omega)}(\partial_{-}\varphi)^{3}\partial_{+}\varphi,
\end{split}
\end{equation}
whereas the $+-$ or $-+$ component of metric varying field equations is given by
\begin{equation}\label{+- component of field eqns}
\partial_{+}\partial_{-}\varphi = \frac{1}{2}\lambda^{2}e^{2\omega} + 2\partial_{+}\varphi
\partial_{-}\varphi + 8G\Gamma e^{2(\varphi - \omega)}(\partial_{+}\varphi\partial_{-}\varphi)
^{2}.
\end{equation}
On the other hand, using the constraint equations (\ref{constraint eqns}), we obtain the
following equation which is associated with field equation coming from the dilaton field
variation
\begin{equation}\label{phi variation eqn}
\begin{split}
\partial_{+}\partial_{-}\omega & = \frac{1}{2}\lambda^{2}e^{2\omega} + 2\partial_{+}\varphi
\partial_{-}\varphi(1 + 8\Gamma G\lambda^{2}e^{2\varphi})\\
 & + 80\Gamma Ge^{2(\varphi - \omega)}(\partial_{+}\varphi\partial_{-}\varphi)^{2}.
\end{split}
\end{equation}
In terms of dimensionless null coordinates $x^{\pm}\rightarrow \lambda x^{\pm}$, the above-mentioned
four equations reduce to the following form
\begin{equation}
\begin{split}
\partial_{+}^{2}\varphi & = 2\partial_{+}\omega\partial_{+}\varphi - 2 \epsilon_1 
e^{2(\varphi - \omega)}(\partial_{+}\varphi)^{3}\partial_{-}\varphi\\
\partial_{-}^{2}\varphi & = 2\partial_{-}\omega\partial_{-}\varphi - 2 \epsilon_1 
e^{2(\varphi - \omega)}(\partial_{-}\varphi)^{3}\partial_{+}\varphi\\
\partial_{+}\partial_{-}\varphi & = \frac{1}{2}e^{2\omega} + 2\partial_{+}\varphi
\partial_{-}\varphi + \epsilon_1 e^{2(\varphi - \omega)}(\partial_{+}\varphi\partial_{-}
\varphi)^{2}\\
\partial_{+}\partial_{-}\omega & = \frac{1}{2}e^{2\omega} + 2\partial_{+}\varphi\partial_{-}
\varphi(1 + \epsilon_1 e^{2\varphi})\\
 & + 10 \epsilon_1 e^{2(\varphi - \omega)}(\partial_{+}\varphi\partial_{-}\varphi)^{2},
\end{split}
\end{equation}
where $\epsilon_1 = 8 \Gamma G\lambda^{2}$. The dimensionless null coordinates are particularly
important for obtaining numerical solutions since $\varphi$ and $\omega$ are dimensionless
quantities.

\subsection{Model 2}

Using the relations in (\ref{useful relations}), we obtain the following constraint equations 
which are the $++$ and $--$ component of metric varying field equations
\begin{equation}\label{constraint eqns Model 2}
\begin{split}
\partial_{+}^{2}\varphi & = 2\partial_{+}\omega\partial_{+}\varphi - 16G\bar{\Gamma} 
e^{ - 2\omega}(\partial_{+}\varphi)^{3}\partial_{-}\varphi\\
\partial_{-}^{2}\varphi & = 2\partial_{-}\omega\partial_{-}\varphi - 16G\bar{\Gamma} 
e^{ - 2\omega}(\partial_{-}\varphi)^{3}\partial_{+}\varphi,
\end{split}
\end{equation}
whereas the $+-$ or $-+$ component of metric varying field equations is given by
\begin{equation}\label{+- component of field eqns Model 2}
\partial_{+}\partial_{-}\varphi = \frac{1}{2}\lambda^{2}e^{2\omega} + 2\partial_{+}\varphi
\partial_{-}\varphi + 8G\bar{\Gamma}e^{ - 2\omega}(\partial_{+}\varphi\partial_{-}\varphi)
^{2}.
\end{equation}
On the other hand, using the constraint equations (\ref{constraint eqns Model 2}), we obtain 
the following equation which is associated with field equation coming from the dilaton field
variation
\begin{equation}\label{phi variation eqn Model 2}
\begin{split}
\partial_{+}\partial_{-}\omega & = \frac{1}{2}\lambda^{2}e^{2\omega} + 2\partial_{+}\varphi
\partial_{-}\varphi(1 + 8\bar{\Gamma} G\lambda^{2}e^{2\varphi})\\
 & + 56\bar{\Gamma} Ge^{ - 2\omega}(\partial_{+}\varphi\partial_{-}\varphi)^{2}.
\end{split}
\end{equation}
In terms of dimensionless null coordinates $x^{\pm}\rightarrow \lambda x^{\pm}$, the above-mentioned
four equations reduce to the following form
\begin{equation}
\begin{split}
\partial_{+}^{2}\varphi & = 2\partial_{+}\omega\partial_{+}\varphi - 2 \epsilon_2 
e^{ - 2\omega}(\partial_{+}\varphi)^{3}\partial_{-}\varphi\\
\partial_{-}^{2}\varphi & = 2\partial_{-}\omega\partial_{-}\varphi - 2 \epsilon_2 
e^{ - 2\omega}(\partial_{-}\varphi)^{3}\partial_{+}\varphi\\
\partial_{+}\partial_{-}\varphi & = \frac{1}{2}e^{2\omega} + 2\partial_{+}\varphi
\partial_{-}\varphi +\epsilon_2 e^{ - 2\omega}(\partial_{+}\varphi\partial_{-}
\varphi)^{2}\\
\partial_{+}\partial_{-}\omega & = \frac{1}{2}e^{2\omega} + 2\partial_{+}\varphi\partial_{-}
\varphi(1 + \epsilon_2)\\
 & + 7 \epsilon_2 e^{ - 2\omega}(\partial_{+}\varphi\partial_{-}\varphi)^{2},
\end{split}
\end{equation}
where $\epsilon_2 = \bar{\Gamma}G\lambda^{2}$.

\section{Corrected CGHS models in $\sigma_{\pm}$ coordinates}
\label{app:Numerical}

In order to obtain the numerical solutions of the field equations, we introduce the new
set of null coordinates $\sigma^{\pm}$ in the following manner
\begin{equation}
\lambda x^{\pm} \rightarrow \pm e^{\pm\lambda\sigma^{\pm}},
\end{equation}
which preserve our conformal gauge choice. Here $x^{\pm}$ and $\sigma^{\pm}$ are the 
dimensionful coordinates. Since we are interested in finding out the solutions which
are time-independent, we define the spacelike coordinate $\sigma = (\sigma^{+} -
\sigma^{-})/2$. Now we introduce the dimensionless coordinates through the mapping 
$x^{\pm}\rightarrow\lambda x^{\pm}$ and $\sigma\rightarrow\lambda\sigma$. Therefore,
in order to find the spatial configurations or solutions of the field equations, we
can substitute $\partial_{\pm} = \pm\frac{1}{2}\partial_{\sigma}$ in the field equations
obtained in the previous section. However, in order to find the numerical solutions 
of these coupled second order differential equation, we introduce the following quantity
\begin{equation}
Y = \frac{1}{4}\nabla_{\mu}\varphi\nabla^{\mu}\varphi = - e^{-2\omega}\partial_{+}\varphi
\partial_{-}\varphi,
\end{equation} 
which satisfy the following differential equations
\begin{equation}
\begin{split}
\partial_{\pm}Y & = - \partial_{\pm}\varphi\left(\frac{1}{2} - 2Y - 8\epsilon e^{2\varphi}
Y^{2}\right)\\
\implies \partial_{\sigma}Y & = - \partial_{\sigma}\varphi\left(\frac{1}{2} - 2Y - 8\epsilon 
e^{2\varphi}Y^{2}\right),
\end{split}
\end{equation}
where we used the field equations in $\varphi$ for Model 1. On the other hand, we also obtain 
the following relation from the definition of $Y$
\begin{equation}
(\partial_{\sigma}\varphi)^{2} = 4Ye^{2\omega} \implies \partial_{\sigma}\varphi = \pm 2
e^{\omega}\sqrt{Y}.
\end{equation} 
However, in the above equation we choose the negative sign. On the other hand, the field equation
in $\omega$ can be expressed as
\begin{equation}
\partial_{\sigma}^{2}\omega = - 4 e^{2\omega}\left(\frac{1}{2} - 2Y(1 + 8\epsilon e^{2\varphi})
 + 80\epsilon e^{2\varphi} Y^{2}\right).
\end{equation}
Therefore, now we have a system of coupled ordinary differential equations in variables $\varphi, 
Y$, and $\omega$ which can be solved numerically given an initial condition. The same set of 
equations in Model 2 can be expressed as
\begin{equation}
\begin{split}
\partial_{\sigma}Y & = - \partial_{\sigma}\varphi\left(\frac{1}{2} - 2Y - 8\epsilon Y^{2}\right),
 \ \partial_{\sigma}\varphi = - 2e^{\omega}\sqrt{Y}\\
\partial_{\sigma}^{2}\omega & = - 4e^{2\omega}\left(\frac{1}{2} - 2Y(1 + 8\epsilon) + 56\epsilon
Y^{2}\right).
\end{split}
\end{equation}
Here we would like to point out that $Y$ is related to $X_{\varphi}$ via the relation $Y = - 
X_{\varphi}/2$.

\section{Hawking flux computation}\label{Appendix I}

\subsection{CGHS model}
In this section, we consider the CGHS model with minimally coupled massless scalar field theory, described by the following action
\begin{equation}
S = \frac{1}{4\pi}\int\sqrt{-g}d^{2}x \Big[e^{-2\varphi}[\mathcal{R} + 4\nabla_{\mu}\varphi
\nabla^{\mu}\varphi + 4\lambda^{2}] - \nabla_{\mu}f\nabla^{\mu}f\Big].
\end{equation}
Considering a matter $f$-showckwave described by the relation $\partial_{+}f\partial_{+}f = 2M/(\lambda x_{0}^{+})\delta(x^{+} - x_{0}^{+})$, the solution of metric and dilaton field are given by the following 
\begin{equation}
e^{-2\varphi} = e^{-2\omega} = - \lambda^{2}x^{+}x^{-} + \frac{M}{\lambda}\left(1 - \frac{x^{+}}{x_{0}^{+}}\right)\theta(x^{+} - x_{0}^{+}).
\end{equation}
We may note that for $x^{+} < x_{0}^{+}$, the above describes the linear dilaton vacuum whereas for $x^{+} > x_{0}^{+}$, it describes a black hole solution of mass $M$, and these two solutions are joined by the $f$-shockwave. Further, both for $x > x_{0}^{+}$ and $x < x_{0}^{+}$, the energy-momentum tensor
vanishes identically for a $f$-shockwave. However, after doing a one-loop computation, it can be shown that the trace of vacuum expectation value of the energy-momentum tensor is given by
\begin{equation}
g^{\mu\nu}\langle T_{\mu\nu}^{f}\rangle = - 4e^{-2\omega}\langle T_{+-}^{f}\rangle = 
\frac{1}{24\pi}\mathcal{R},
\end{equation}
which leads to the following expression of $\langle T_{+-}^{f}\rangle$
\begin{equation}
\langle T_{+-}^{f}\rangle = - \frac{e^{2\omega}}{96\pi}\mathcal{R} = - \frac{1}{12\pi}
\partial_{+}\partial_{-}\omega.
\end{equation}
Now we look at the conservation of energy-momentum tensor which is given by $\nabla^{\mu}
T_{\mu\nu} = 0$. For $\nu = +$, we obtain the following relation
\begin{equation}
\begin{split}
\nabla_{-} & \langle T_{++}^{f}\rangle + \nabla_{+}\langle T_{-+}^{f}\rangle = 0\\
\implies \partial_{-}\langle T_{++}^{f}\rangle & = - (\partial_{+} - 2\partial_{+}\omega)
\langle T_{-+}^{f}\rangle\\
 & = \frac{1}{12\pi}\left(\partial_{+}^{2}\partial_{-}\omega - 2\partial_{+}\omega
 \partial_{+}\partial_{-}\omega\right)\\
 & = \frac{1}{12\pi}\partial_{-}\left(\partial_{+}^{2}\omega - (\partial_{+}\omega)^{2}
 \right),
\end{split}
\end{equation}
which implies
\begin{equation}
\langle T_{++}^{f}\rangle = \frac{1}{12\pi}\left(\partial_{+}^{2}\omega - (\partial_{+}\omega)^{2}
 + t_{+}(x^{+})\right).
\end{equation}
In the similar manner, considering $\nu = -$ component of conservation equation, we are
able to show that
\begin{equation}
\langle T_{--}^{f}\rangle = \frac{1}{12\pi}\left(\partial_{-}^{2}\omega - (\partial_{-}\omega)^{2}
 + t_{-}(x^{-})\right).
\end{equation}
In the above expressions of $T_{\pm\pm}^{f}$, $t_{\pm}(x^{\pm})$ can be obtained from the
boundary conditions. In terms of the following null coordinates, 
\begin{equation}
\begin{split}
\lambda x^{+} & = e^{\lambda\sigma^{+}}, \ e^{ - \lambda\sigma^{-}} = - \lambda x^{-} - 
\frac{\bar{M}}{\lambda x_{0}^{+}}\\
\end{split}
\end{equation}
we may write the following relations
\begin{equation}
\begin{split}
dx^{+}dx^{-} & = e^{\lambda(\sigma^{+} - \sigma^{-})}d\sigma^{+}d\sigma^{-}\\
\xi & = \lambda^{2}x^{+}x^{-} = - e^{\lambda(\sigma^{+} - \sigma^{-})} - \frac{\bar{M}}
{\lambda x_{0}^{+}}e^{\lambda\sigma^{+}}\\
\implies z & = \begin{cases}
\frac{1}{e^{\lambda(\sigma^{+} - \sigma^{-})} + \frac{\bar{M}}
{\lambda x_{0}^{+}}e^{\lambda\sigma^{+}}}, \ \sigma^{+} < \sigma_{0}^{+}\\
\frac{1}{e^{\lambda(\sigma^{+} - \sigma^{-})} + \frac{\bar{M}}{\lambda x_{0}^{+}}e^{\lambda
\sigma_{0}^{+}}}, \ \sigma^{+} > \sigma_{0}^{+} 
\end{cases}
\end{split}
\end{equation}
which implies the CGHS metric element reduces to 
\begin{equation}
e^{2\bar{\omega}} = \begin{cases}
\left(1 + \frac{\bar{M}}{\lambda x_{0}^{+}}e^{\lambda\sigma^{-}}\right)^{-1}, & \ \sigma^{+}
< \sigma_{0}^{+}\\
\left(1 + \frac{\bar{M}}{\lambda x_{0}^{+}}e^{\lambda(\sigma^{-} - \sigma^{+} + \sigma_{0}^{+}
)}\right)^{-1}, & \ \sigma^{+} > \sigma_{0}^{+}.
\end{cases}
\end{equation}
For collapsing $f$-wave, the components of $T^{f}$ should vanish identically in the linear
dilaton region, and there should not be any incoming radiation along $\mathcal{J}_{R}^{-}$
except for the classical $f$-wave at $\sigma_{0}^{+}$. Therefore, for $\sigma^{+}\rightarrow
-\infty$, we find
\begin{equation}\label{leading order boundary terms}
t_{+}^{(0)} = 0, \ t_{-}^{(0)} = \frac{\lambda^{2}}{4}\left( 1 - \frac{1}{\left(1 + 
\frac{\bar{M}}{\lambda x_{0}^{+}}e^{\lambda\sigma^{-}}\right)^{2}}\right)
\end{equation} 
if we consider the $f$-shockwave metric of the CGHS model. This finally leads us to the Hawking
flux at $\mathcal{J}_{R}^{+}$ by taking the limit $\sigma^{-}\rightarrow\infty$
\begin{equation}
\langle T_{--}^{f}\rangle = \frac{\lambda^{2}}{48}.
\end{equation}
\subsection{Model 1}
\label{Appendix F}
%
In this section, we consider CGHS model with corrections in which a massless scalar field
is minimally coupled to the background spacetime. Therefore, the action we consider is of 
the following form
\begin{equation}
\begin{split}
S & = \Lambda^{2}(\bar{S}_{0} + \bar{S}_{\text{corr}}) + S_{m}\\
\bar{S}_{0} & = \int\sqrt{-g}d^{2}x \ \frac{e^{-2\varphi}}{4G}\left(\mathcal{R} + 
4\nabla_{\mu}\varphi\nabla^{\mu}\varphi + 4\lambda^{2}\right)\\
\bar{S}_{\text{corr}} & = - \Gamma\int\sqrt{-g}d^{2}x \ (\nabla_{\mu}\varphi\nabla^{\mu}
\varphi)^{2}\\
S_{m} & = - \frac{1}{4\pi}\int\sqrt{-g}d^{2}x \ g^{\mu\nu}\nabla_{\mu}f\nabla_{\nu}f, 
\end{split}
\end{equation}
where the metric $g_{\mu\nu}$ is considered to be the reduced two-dimensional spacetime
metric. The field equation of matter is given by the massless Klein-Gordon equation
\begin{equation}
\Box f = 0 \implies \partial_{+}\partial_{-}f = 0.
\end{equation}
In this section, we consider the general linearised field equations in null coordinates
$\sigma^{\pm}$ defined in the following manner
\begin{equation}
\lambda x^{+} = e^{\lambda\sigma^{+}}, \ e^{ - \lambda\sigma^{-}} = - \lambda x^{-} - 
\frac{\bar{M}}{\lambda x_{0}^{+}},
\end{equation} 
where $x_{0}^{+}$ is a constant. We may note that in this coordinate choice $\bar{\omega}
\neq \bar{\varphi}$ where
\begin{equation}
e^{2\bar{\omega}} = \begin{cases}
\left(1 + \frac{\bar{M}}{\lambda x_{0}^{+}}e^{\lambda\sigma^{-}}\right)^{-1}, & \ \sigma^{+}
< \sigma_{0}^{+}\\
\left(1 + \frac{\bar{M}}{\lambda x_{0}^{+}}e^{\lambda(\sigma^{-} - \sigma^{+} + \sigma_{0}^{+}
)}\right)^{-1}, & \ \sigma^{+} > \sigma_{0}^{+},
\end{cases}
\end{equation}
and 
\begin{equation}
e^{2\bar{\varphi}} = \begin{cases}
\frac{1}{e^{\lambda(\sigma^{+} - \sigma^{-})} + \frac{\bar{M}}{\lambda x_{0}^{+}}
e^{\lambda\sigma^{+}}}, & \ \sigma^{+} < \sigma_{0}^{+}\\
\frac{1}{e^{\lambda(\sigma^{+} - \sigma^{-})} + \frac{\bar{M}}{\lambda x_{0}^{+}}
e^{\lambda\sigma_{0}^{+}}}, & \ \sigma^{+} > \sigma_{0}^{+}
\end{cases}
\end{equation}
If we choose to work dimensionless null coordinates $\lambda\sigma^{\pm}$, then the constraint 
equations and the equations of motion can be expressed as
\begin{equation}
\begin{split}
\partial_{+}^{2}\varphi & = 2\partial_{+}\omega\partial_{+}\varphi - 16\epsilon 
e^{2(\varphi - \omega)}(\partial_{+}\varphi)^{3}\partial_{-}\varphi\\
\partial_{-}^{2}\varphi & = 2\partial_{-}\omega\partial_{-}\varphi - 16\epsilon 
e^{2(\varphi - \omega)}(\partial_{-}\varphi)^{3}\partial_{+}\varphi\\
\partial_{+}\partial_{-}\varphi & = \frac{1}{2}e^{2\omega} + 2\partial_{+}\varphi
\partial_{-}\varphi + 8\epsilon e^{2(\varphi - \omega)}(\partial_{+}\varphi\partial_{-}\varphi)
^{2}\\
\partial_{+}\partial_{-}\omega & = \frac{1}{2}e^{2\omega} + 2\partial_{+}\varphi
\partial_{-}\varphi(1 + 8\epsilon e^{2\varphi})\\
 & + 80\epsilon e^{2(\varphi - \omega)}(\partial_{+}\varphi\partial_{-}\varphi)^{2},
\end{split}
\end{equation}
where $\epsilon = G\Gamma\lambda^{2}$. If we consider the perturbations of the following form
\begin{equation}
\omega = \bar{\omega} + \epsilon\delta\omega, \ \varphi = \bar{\varphi} + \epsilon\delta\varphi,
\end{equation}
then we obtain the following linearised equations
\begin{equation}
\begin{split}
\partial_{+}^{2}\delta\varphi & = 2\partial_{+}\delta\omega\partial_{+}\bar{\varphi} + 2
\partial_{+}\bar{\omega}\partial_{+}\delta\varphi\\
 & - 16e^{2(\bar{\varphi} - \bar{\omega})}(\partial_{+}\bar{\varphi})^{3}\partial_{-}
 \bar{\varphi}\\
\partial_{-}^{2}\delta\varphi & = 2\partial_{-}\delta\omega\partial_{-}\bar{\varphi} + 2
\partial_{-}\bar{\omega}\partial_{-}\delta\varphi\\
 & - 16e^{2(\bar{\varphi} - \bar{\omega})}(\partial_{-}\bar{\varphi})^{3}\partial_{+}
 \bar{\varphi}\\
\partial_{+}\partial_{-}\delta\varphi & = e^{2\bar{\omega}}\delta\omega + 2\partial_{+}
\bar{\varphi}\partial_{-}\delta\varphi + 2\partial_{+}\delta\varphi\partial_{-}\bar{\varphi}
\\
 & + 8e^{2(\bar{\varphi} - \bar{\omega})}(\partial_{+}\bar{\varphi}\partial_{-}\bar{\varphi})
^{2}\\
\partial_{+}\partial_{-}\delta\omega & = e^{2\bar{\omega}}\delta\omega + 2\partial_{+}
\bar{\varphi}\partial_{-}\delta\varphi + 2\partial_{+}\delta\varphi\partial_{-}\bar{\varphi}
\\
 & + 16e^{2\bar{\varphi}}\partial_{+}\bar{\varphi}\partial_{-}\bar{\varphi} + 80e^{2(\bar{\varphi}
 - \bar{\omega})}(\partial_{+}\bar{\varphi}\partial_{-}\bar{\varphi})^{2}.
\end{split}
\end{equation}
Since for $\sigma^{+} < \sigma_{0}^{+}$, we expect the expressions of $\langle T_{\pm\pm}^{f}\rangle$ to vanish which fix the expressions of $t_{\pm}$, and they are given by
\begin{equation}\label{boundary terms}
\begin{split}
t_{\pm}(\sigma^{\pm}) & = (\partial_{\pm}\omega)^{2} - \partial_{\pm}^{2}\omega = (\partial_{\pm}\bar{\omega})^{2} - \partial_{\pm}^{2}\bar{\omega}\\
 & + 2\epsilon\partial_{\pm}\bar{\omega}\partial_{\pm}\delta\omega - \epsilon\partial_{\pm}^{2}
 \delta\omega\\
 & = t_{\pm}^{(0)} + \epsilon(2\partial_{\pm}\bar{\omega}\partial_{\pm}\delta\omega - \partial_{\pm}^{2}
 \delta\omega),
\end{split}
\end{equation}
where the expressions of $t_{0}^{\pm}$ are known to us exactly from the CGHS model. 

We may first note that, for $\sigma^{+} > \sigma_{0}^{+}$, the metric field $\bar{\omega}$ and 
the dilaton field both depend on the combination $\lambda(\sigma^{+} - \sigma^{-})$. Therefore, considering the above symmetry for the linearised solutions, we replace the derivative operators 
by $\partial_{+} = - \partial_{-} = \partial$. This leads to the following linearised equations
\begin{equation}\label{linearised eqns. second null}
\begin{split}
\partial^{2}\delta\varphi & = 2\partial\delta\omega\partial\bar{\varphi} + 2\partial\bar{\omega}
\partial\delta\varphi + 16e^{2(\bar{\varphi} - \bar{\omega})}(\partial\bar{\varphi})^{4}\\
\partial^{2}\delta\varphi & = - e^{2\bar{\omega}}\delta\omega + 4\partial\bar{\varphi}\partial
\delta\varphi - 8e^{2(\bar{\varphi} - \bar{\omega})}(\partial\bar{\varphi})^{4}\\
\partial^{2}\delta\omega & = - e^{2\bar{\omega}}\delta\omega + 4\partial\bar{\varphi}\partial
\delta\varphi + 16e^{2\bar{\varphi}}(\partial\bar{\varphi})^{2} - 80e^{2(\bar{\varphi} - 
\bar{\omega})}(\partial\bar{\varphi})^{4}.
\end{split}
\end{equation}
Comparing the first two equations, we obtain the following relation
\begin{equation}
e^{2\bar{\omega}}\delta\omega + 2\partial\delta\omega\partial\bar{\varphi} + 2\partial\delta
\varphi(\partial\bar{\omega} - 2\partial\bar{\varphi}) + 24e^{2(\bar{\varphi} - \bar{\omega})}
(\partial\bar{\varphi})^{4} = 0.
\end{equation}
On the other hand, neglecting $\partial^{2}\delta\omega$, we obtain the following relation
\begin{equation}
0 = - e^{2\bar{\omega}}\delta\omega + 4\partial\bar{\varphi}\partial\delta\varphi + 
16e^{2\bar{\varphi}}(\partial\bar{\varphi})^{2} - 80e^{2(\bar{\varphi} - \bar{\omega})}
(\partial\bar{\varphi})^{4}
\end{equation}
Therefore, we obtain the two-coupled first order differential equations in $\delta\varphi, \delta
\omega$  which can principle be solved numerically. However, in the limit $\sigma^{+} \rightarrow
\infty$, we obtain the following relations
\begin{equation}
\begin{split}
e^{2\bar{\omega}} & \rightarrow 1, \ e^{2\bar{\varphi}} \rightarrow 0\\
\partial\bar{\omega} & \rightarrow 0, \ \partial\bar{\varphi} \rightarrow - \frac{1}{2}.
\end{split}
\end{equation}
which makes the coupled differential equations in this limit as
\begin{equation}
\delta\omega - \partial\delta\omega + 2\partial\delta\varphi = 0, \ \delta\omega + 
2\partial\delta\varphi = 0 \implies \partial\delta\omega = 0.
\end{equation}
Therefore, we can choose the $\delta\omega = 0$ in the limit $\sigma^{+}\rightarrow\infty$.

On the other hand, in $\sigma^{+} < \sigma_{0}^{+}$ limit, we have the following generalized 
relations
\begin{equation}
\begin{split}
\partial_{+}^{2}\delta\varphi & = 2\partial_{+}\delta\omega\partial_{+}\bar{\varphi} + 2
\partial_{+}\bar{\omega}\partial_{+}\delta\varphi\\
 & - 16e^{2(\bar{\varphi} - \bar{\omega})}(\partial_{+}\bar{\varphi})^{3}\partial_{-}
 \bar{\varphi}\\
\partial_{-}^{2}\delta\varphi & = 2\partial_{-}\delta\omega\partial_{-}\bar{\varphi} + 2
\partial_{-}\bar{\omega}\partial_{-}\delta\varphi\\
 & - 16e^{2(\bar{\varphi} - \bar{\omega})}(\partial_{-}\bar{\varphi})^{3}\partial_{+}
 \bar{\varphi}\\
\partial_{+}\partial_{-}\delta\varphi & = e^{2\bar{\omega}}\delta\omega + 2\partial_{+}
\bar{\varphi}\partial_{-}\delta\varphi + 2\partial_{+}\delta\varphi\partial_{-}\bar{\varphi}
\\
 & + 8e^{2(\bar{\varphi} - \bar{\omega})}(\partial_{+}\bar{\varphi}\partial_{-}\bar{\varphi})
^{2}\\
\partial_{+}\partial_{-}\delta\omega & = e^{2\bar{\omega}}\delta\omega + 2\partial_{+}
\bar{\varphi}\partial_{-}\delta\varphi + 2\partial_{+}\delta\varphi\partial_{-}\bar{\varphi}
\\
 & + 16e^{2\bar{\varphi}}\partial_{+}\bar{\varphi}\partial_{-}\bar{\varphi} + 80e^{2(\bar{\varphi}
 - \bar{\omega})}(\partial_{+}\bar{\varphi}\partial_{-}\bar{\varphi})^{2}.
\end{split}
\end{equation}
Moreover, we also have the following relations
\begin{equation}
\begin{split}
\partial_{+}\bar{\omega} & = 0, \ \partial_{-}\bar{\omega} = \frac{1}{2}(e^{2\bar{\omega}} 
- 1)\\
\partial_{+}\bar{\varphi} & = - \frac{1}{2}, \ \partial_{-}\bar{\varphi} = \frac{1}{2}e^{2
\bar{\omega}}, \ e^{2\bar{\varphi}} = e^{\lambda(\sigma^{-} - \sigma^{+})}e^{2\bar{\omega}},
\end{split}
\end{equation}
for $\sigma^{+} < \sigma_{0}^{+}$ which leads to the following linearised equations
\begin{equation}
\begin{split}
\partial_{+}^{2}\delta\varphi & = - \partial_{+}\delta\omega + e^{2\bar{\varphi}}\\
\partial_{-}^{2}\delta\varphi & = \partial_{-}\delta\omega e^{2\bar{\omega}} + 
(e^{2\bar{\omega}} - 1)\partial_{-}\delta\varphi + e^{2(\bar{\varphi} + 2\bar{\omega})}\\
\partial_{+}\partial_{-}\delta\varphi & = e^{2\bar{\omega}}\delta\omega - \partial_{-}
\delta\varphi + e^{2\bar{\omega}}\partial_{+}\delta\varphi + \frac{1}{2}e^{2(\bar{\varphi} 
+ \bar{\omega})}\\
\partial_{+}\partial_{-}\delta\omega & = e^{2\bar{\omega}}\delta\omega - \partial_{-}
\delta\varphi + e^{2\bar{\omega}}\partial_{+}\delta\varphi + e^{2(\bar{\varphi} + \bar{\omega})}
\end{split}
\end{equation}
We may note that in $\sigma^{+}<\sigma_{0}^{+}$ domain, the expressions in (\ref{boundary terms})
reduce to the following form
\begin{equation}
\begin{split}
t_{+} & = t_{+}^{(0)} - \epsilon\partial_{+}^{2}\delta\omega =  - \epsilon\partial_{+}^{2}
\delta\omega\\
t_{-} & = t_{-}^{(0)} + \epsilon[(e^{2\bar{\omega}} - 1)\partial_{-}\delta\omega - \partial_{-}^{2}
\delta\omega],
\end{split}
\end{equation}
where $t_{0}^{-}$ is given in (\ref{leading order boundary terms}). As a result, we obtain the
following formal expressions of $\langle T_{++}^{f}\rangle$ and $\langle T_{--}^{f}\rangle$ in 
$\sigma^{+}\rightarrow\infty$
\begin{equation}
\begin{split}
\langle T_{++}^{f}\rangle & = \frac{1}{12\pi}\Big[\partial_{+}^{2}\bar{\omega} - (\partial_{+}
\bar{\omega})^{2} + \epsilon[\partial_{+}^{2}\delta\omega - 2\partial_{+}\bar{\omega}\partial
_{+}\delta\omega](\sigma^{+}\rightarrow\infty)\\
 & - \epsilon\partial_{+}^{2}\delta\omega(\sigma^{+}\rightarrow - \infty)\Big]\\
 & = \frac{\epsilon}{12\pi}\Big[\partial_{+}^{2}\delta\omega(\sigma^{+}\rightarrow\infty)
 - \partial_{+}^{2}\delta\omega(\sigma^{+}\rightarrow - \infty)\Big]\\
\langle T_{--}^{f}\rangle & = \frac{1}{12\pi}\Big[\partial_{-}^{2}\bar{\omega} - (\partial_{-}
\bar{\omega})^{2} + \epsilon[\partial_{-}^{2}\delta\omega - 2\partial_{-}\bar{\omega}\partial
_{-}\delta\omega](\sigma^{+}\rightarrow\infty)\\
 & + t_{-}^{0} + \epsilon[(e^{2\bar{\omega}} - 1)\partial_{-}\delta\omega - \partial_{-}^{2}
\delta\omega](\sigma^{+}\rightarrow -\infty)\Big]\\
 & = \frac{1}{12\pi}\Big[t_{-}^{(0)} + \epsilon[\partial_{-}^{2}\delta\omega(\sigma^{+}\rightarrow\infty)
 - \partial_{-}^{2}\delta\omega(\sigma^{+}\rightarrow - \infty)]\\
 & + \epsilon(e^{2\bar{\omega}} - 1)\partial_{-}\delta\omega(\sigma^{+}\rightarrow -\infty)\Big], 
\end{split}
\end{equation}
using the dimensionfull null coordinates $\sigma^{\pm}$.

\subsection{Model 2}

In Model 2, we obtain the following set of linearised field equations in terms of dimensionless null coordinates
\begin{equation}
\begin{split}
\partial_{+}^{2}\delta\varphi & = 2\partial_{+}\bar{\omega}\partial_{+}\delta\varphi + 2\partial_{+}\delta\omega\partial_{+}\bar{\varphi} - 16e^{-2\bar{\omega}}(\partial_{+}\bar{\varphi})^{3}\partial_{-}\bar{\varphi}\\
\partial_{-}^{2}\delta\varphi & = 2\partial_{-}\bar{\omega}\partial_{-}\delta\varphi + 2\partial_{-}\delta\omega\partial_{-}\bar{\varphi} - 16e^{-2\bar{\omega}}(\partial_{-}\bar{\varphi})^{3}\partial_{+}\bar{\varphi}\\
\partial_{+}\partial_{-}\delta\varphi & = 2\partial_{+}\bar{\varphi}\partial_{-}\delta\varphi + 2\partial_{+}\delta\varphi\partial_{-}\bar{\varphi} + e^{2\bar{\omega}}\delta\omega\\
 & + 8e^{-2\bar{\omega}}(\partial_{+}\bar{\varphi}\partial_{-}\bar{\varphi})^{2}\\
\partial_{+}\partial_{-}\delta\omega & = e^{2\bar{\omega}}\delta\omega + 2\partial_{+}\delta\varphi\partial_{-}\bar{\varphi} + 2\partial_{+}\bar{\varphi}\partial_{-}\delta\varphi\\
 & + 16\partial_{+}\bar{\varphi}\partial_{-}\bar{\varphi} + 56e^{-2\bar{\omega}}(\partial_{+}\bar{\varphi}\partial_{-}\bar{\varphi})^{2}. 
\end{split}
\end{equation}
Following the previous subsection, we may write the above equations as
\begin{equation}
\begin{split}
\partial^{2}\delta\varphi & = 2\partial\bar{\omega}\partial\delta\varphi + 2\partial\delta\omega\partial\bar{\varphi} + 16e^{-2\bar{\omega}}(\partial\bar{\varphi})^{4}\\
\partial^{2}\delta\varphi & = 4\partial\bar{\varphi}\partial\delta\varphi - e^{2\bar{\omega}}\delta\omega - 8e^{-2\bar{\omega}}(\partial\bar{\varphi})^{4}\\
\partial^{2}\delta\omega & = 4\partial\bar{\varphi}\partial\delta\varphi - e^{2\bar{\omega}}\delta\omega + 16(\partial\bar{\varphi})^{2} - 56e^{-2\bar{\omega}}(\partial\bar{\varphi})^{4},
\end{split}
\end{equation}
in the domain $\sigma^{+} > \sigma_{0}^{+}$. On the other hand, in the domain $\sigma^{+} < \sigma_{0}^{+}$, we obtain the following set of coupled differential equations
\begin{equation}
\begin{split}
\partial_{+}^{2}\delta\varphi & = 1 - \partial_{+}\delta\omega\\
\partial_{-}^{2}\delta\varphi & = (e^{2\bar{\omega}} - 1)\partial_{-}\delta\varphi + e^{2\bar{\omega}}
\partial_{-}\delta\omega + e^{4\bar{\omega}}\\
\partial_{+}\partial_{-}\delta\varphi & = - \partial_{-}\delta\varphi + e^{2\bar{\omega}}\partial_{+}\delta\varphi + e^{2\bar{\omega}}\delta\omega + \frac{1}{2}e^{2\bar{\omega}}\\
\partial_{+}\partial_{-}\delta\omega & = e^{2\bar{\omega}}\delta\omega + e^{2\bar{\omega}}\partial_{+}\delta\varphi - \partial_{-}\delta\varphi - \frac{1}{2}e^{2\bar{\omega}}.
\end{split}
\end{equation}

Like in Model 1, we can obtain the Hawking flux from the above expression.

\end{document}